# Statistical reproducibility of meta-analysis research claims for medical mask use in community settings to prevent COVID infection

S. Stanley Young[1] and Warren B. Kindzierski[2]

[1] CGStat, Raleigh, NC, USA
[2] Independent consultant, St Albert, Alberta, Canada

Correspondence: Warren B. Kindzierski, 12 Hart Place, St Albert, Alberta, T8N 5R1, Canada. Email: wbk@shaw.ca or warrenk@ualberta.ca.

**ABSTRACT** (word count=216)
The coronavirus pandemic (COVID) has been an exceptional test of current scientific evidence that inform and shape policy. Many US states, cities, and counties implemented public orders for mask use on the notion that this intervention would delay and flatten the epidemic peak and largely benefit public health outcomes. P-value plotting was used to evaluate statistical reproducibility of meta-analysis research claims of a benefit for medical (surgical) mask use in community settings to prevent COVID infection. Eight studies (seven meta-analyses, one systematic review) published between 1 January 2020 and 7 December 2022 were evaluated. Base studies were randomized control trials with outcomes of medical diagnosis or laboratory-confirmed diagnosis of viral (Influenza or COVID) illness. Self-reported viral illness outcomes were excluded because of awareness bias. No evidence was observed for a medical mask use benefit to prevent viral infections in six p-value plots (five meta-analyses and one systematic review). Research claims of no benefit in three meta-analyses and the systematic review were reproduced in p-value plots. Research claims of a benefit in two meta-analyses were not reproduced in p-value plots. Insufficient data were available to construct p-value plots for two meta-analyses because of overreliance on self-reported outcomes. These findings suggest a benefit for medical mask use in community settings to prevent viral, including COVID infection, is unproven.

## INTRODUCTION
### Background
The coronavirus pandemic (COVID) in 2020 has been an exceptional test of current scientific evidence that inform and shape government policy. Governments around the world were faced with a disease whose significance was initially uncertain. Governments acted swiftly given further uncertainties in the capacity of their health care systems to respond to the virus.

The World Health Organization (WHO) declared COVID a pandemic on March 11, 2020 (Lavezzo et al. 2020, Members 2020, CDC 2022). Countless governments followed up with restrictive pandemic policies. Examples of policies imposed as large-scale restrictions on populations included (Gostin et al. 2020, Jenson 2020, Magness 2021): public orders for mask wearing in community settings; quarantine (stay-at-home) orders; nighttime curfews; closing of schools, universities, and many businesses; and bans on large gatherings.

Early in the pandemic, the U.S. Centers for Disease Control and Prevention (CDC) recommended a cautious approach that patients in health care settings under investigation for symptoms of suspected COVID should wear a medical mask as soon as they are identified (Patel & Jernigan 2020).  On April 30, 2020, the CDC recommended that all people wear a mask outside of their home (CDC 2022).

This recommendation came about after emerging data reported transmission of the COVID virus from persons without symptoms and recognition that there was airborne transmission. Initial recommendations were on using cloth face coverings that could be made more widely available in the community than medical masks (Furukawa et al. 2020). This is in spite of Balazy et al. (2006) and Inglesby et al. (2006) reporting that medical masks do little to prevent inhalation of small droplets bearing Influenza virus. Cloth face covering used was intended to preserve personal protective equipment such as medical masks and N95 respirators to highest-risk exposures in health care settings (Furukawa et al. 2020).

Mathematical modelling studies using simulated pandemic scenarios were used to justify durations of restrictions imposed on people, ranging from 2 weeks to months (CDC 2017, Jenson 2020). These restrictions were intended to "flatten the epidemic curve" (Matrajt & Leung 2020). The phrase "flatten the epidemic curve" was originally utilized by the CDC (2007) in pandemic preparation to justify use of nonpharmaceutical interventions and antiviral medications to delay and flatten the epidemic peak. Inglesby et al. (2006) spoke against some of these measures (i.e., quarantine; closing of schools and universities, cancelling or postponing meetings or events involving large gatherings) for the control of pandemic Influenza.

A rationale for flattening the epidemic curve in a pandemic is spreading out health care demands resulting from a high incidence peak that could potentially overwhelm health care utilization capacity (Jenson 2020). Restrictions implemented by governments – including public masking in community settings, however, became lengthy impositions as policy targets developed by public health official shifted (Magness 2021). Political influence dominated both the initiation and ultimate duration of these restrictions in the US (Kosnik & Bellas 2020).

**Research reproducibility**
The overall research capacity response to COVID since the beginning of 2020 has been massive (Kinsella et al. 2020, Chu et al. 2021, Ioannidis et al. 2022). To show the extent of this response, the Advanced Search Builder capabilities of the PubMed search engine was used to approximate the number of COVID publications. The terms covid[Title] OR sars-cov-2[Title] were used for the period 2020-2023 (search performed December 7, 2022). The search returned 250,492 listings in the National Library of Medicine data base.

Prior to the COVID pandemic, researchers have increasingly recognized that only a small portion of published research may be reproducible (e.g., Ioannidis 2005, 2022, Ioannidis et al. 2011, Keown 2012, Begley & Ioannidis 2015, Iqbal et al. 2016, Randall & Welser 2018, Stodden et al. 2018).

Lack of research transparency is a reason for research irreproducibility (Landis et al. 2012), due to biased study designs, flexibility in research practices, low statistical power and chasing statistical significance, and selective analysis and reporting (Kavvoura et al. 2007, Ioannidis 2008, Ioannidis et al. 2011, Ware & Munafo 2015). Whereas, transparent research enables openness of study design, verification of results, creation of new findings with previous knowledge, and effective inquiry of research (Munafo et al. 2017).

Many research studies have been published in response to COVID. However, the reproducibility of some of this research is uncertain, particularly where observational data are used (Bramstedt 2020, Peng & Hicks 2021). The current situation of irreproducible research may be such that not much is different since COVID (Gustot 2020, Sumner et al. 2020, Paez 2021).

**Meta-analysis**

Meta-analysis is a procedure for statistically combining data (test statistics) from multiple studies that address a common research question or claim (Egger et al., 2001). An example of a research question or claim (i.e., cause−effect science claim) addressed in meta-analysis is whether an intervention (or risk factor) is causal of a health outcome.

A meta-analysis assesses a claim by taking a test statistic (e.g., risk ratio, odds ratio, hazard ration, etc.) along with a measure of its reliability (e.g., confidence intervals) from multiple individual intervention—health outcome studies (called base papers) identified the literature. These statistics are combined to give a supposedly more reliable estimate of an effect (Young & Kindzierski, 2019).

It first involves a systematic review. The systematic review of a clearly formulated research question is intended to use systematic and explicit methods to identify, select, and critically appraise relevant research, and to collect and analyze data from the identified studies (Moher et al., 2009). A meta-analysis then selects and then combines test statistics of the identified studies from the systematic review.

One component of replication—performance of another study statistically confirming the same hypothesis or claim—is a cornerstone of science and replication of research claims is important before cause−effect claims can be made (Moonesinghe et al. 2007). However, if a replication study result does not concur with a prevailing paradigm, it might not be published. Also, if a similar faulty method is used in a replication study as in an original study, or if studies with negative findings are not published whereas studies with positive findings are, then a false claim can be taken as fact, canonized (Nissen et al., 2016).

Meta-analysis is ranked high in the medical evidence-based pyramid – above randomized trials, and case–control and cohort studies (Murad et al. 2016, Herner 2019). A major assumption of meta-analysis is that summary statistics drawn from base papers for analysis are unbiased estimates of an effect of interest (Boos & Stefanski 2013). Given these characteristics, independent evaluation of published meta-analysis on a common research question has been used to assess the statistical reproducibility of a claim coming from that field of research (Young & Kindzierski 2019, Kindzierski et al. 2021, Young & Kindzierski 2022, Young & Kindzierski 2023).

Given potentially large data sets available to medical researchers today, intervention−health outcome studies require a strong statistical component to establish informative and interpretable intervention−risk/benefit associations and research claims made from these associations. A statistical approach (p-value plotting after Schweder & Spjøtvoll 1982) was used in this study to evaluate reproducibility of meta-analysis research claims related to benefit of mask use in community settings to prevent COVID infection. The focus was on medical masks of the type shown in Figure 1.

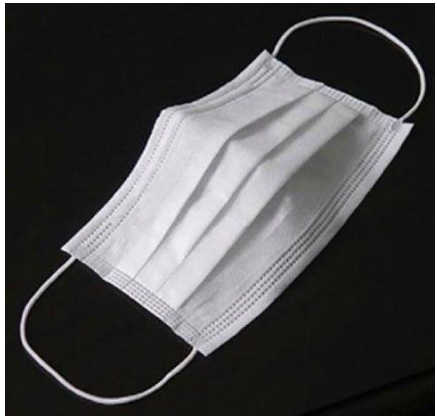

Figure 1 Medical mask.

**METHODS**

We first wanted to show the number of listings of meta-analysis studies cited in literature related to some aspect of COVID. The PubMed search engine was again used. The terms ((covid[Title]) OR (sars-cov-2[Title])) AND (meta-analysis[Title]) [timeline 2020-2023] were used on December 7, 2022. The search returned 3,256 listings in the National Library of Medicine data base. This included 633 listings for 2020, 1,300 listings for 2021, and 1,323 listings thus far for 2022. This is considered an astonishing amount in that a meta-analysis is a summary of available papers.

**Respiratory virus airborne transmission characteristics**

Viruses are one of the smallest known bioaerosols, with a particle diameter ranging from 20 to 300 nm (0.02−0.3 µm) (Balazy et al. 2006). The COVID (sars-cov-2) virus has a reported size range of 60−160 nm (0.06−0.16 µm) (Bar-On et al. 2020, Menter et al. 2020, Zhu et al. 2020). This is similar to the reported size range of Influenza respiratory viruses (80−120 nm, 0.08−0.12 µm) (Stanley 1944, Mosley and Wyckoff 1946, NIH 2017). Whereas Rhinovirus – a virus responsible for an estimated 30−35% of all adult colds during cold and flu season (NIH 2009) – is smaller with a diameter ~30 nm, 0.03 µm (Stott and Killington, 1972).

Regardless of differing virus sizes, most respiratory viruses are transmitted through secretion fluids during breathing in the form of aerosols (<5 µm) or droplets (>5 µm) rather than isolated viruses (Tellier 2006, 2009, Clase et al. 2020, Prather et al., 2020, Meyerowitz et al. 2021, Wang et al. 2021). Both RNA fragments from Influenza and COVID viruses have been detected in aerosols ranging from 0.25 to >4 µm (Wang et al. 2021).

When viral-infected human hosts breathe, talk, eat, cough or sneeze, they emit aerosol particles across a range of sizes (Han et al. 2013, Wang et al. 2021), and respiratory viruses are in those particles (Fennelly 2020, Meyerowitz et al. 2021). For example, aerosol particles respired from simple breathing are small (size range 0.2 to 0.6 μm), and once emitted can be present in an enclosed setting for several hours (Scheuch 2020). Asymptomatic carriers of a virus do not cough and sneeze, and therefore do not expel large respiratory droplets.

Medical masks (e.g., Figure 1) provide protection against large droplets. However, smaller particulates (aerosols) are less effectively filtered. Aerosol particles between ~0.1 and 0.5 μm are not easily filtered out of the surrounding air by any physical mechanism, and there continues to be uncertainties about use of conventional medical masks to separate (remove) these small aerosol particles (Scheuch 2020).

Belkin (1996) states that there are a couple of ways in which virus-laden aerosol particles can contribute to infection of a mask wearer when these particles are present in the breathing zone of the mask wearer. These include (Belkin 1996): aerosol particle penetration through a mask during inhalation, and inhalation of air containing aerosol particles from the side of a mask due to incorrect wear or increased mask resistance or poor string tension.

A mask wearer breathing out moist air increases mask resistance (Belkin 1996). Simple breathing has been shown to release up to 7,200 aerosol particles per liter of exhaled air (Wang et al. 2021). While this can reduce aerosol penetration through the mask, it worsens the problem of inhaling virus-laden aerosols from the side of the mask (Inglesby et al. 2006).

**Study selection**
The randomized controlled trial (RCT) is recognized as a 'gold standard' for assessing the efficacy of an intervention (O'Conner et al. 2008). For this evaluation, interest was in "meta-analysis" or "systematic review" studies of RCTs investigating community medical mask use for prevention of viral infection. The focus in this evaluation was on Influenza and COVID viruses because of their similar size ranges; keeping in mind it is not the virus itself but airborne transmission of aerosols or droplets containing viruses that is important for infection.

Another distinction made in this evaluation is the nature of the outcome for assessing the potential benefit of mask use. Numerous types of outcome measures have been used in mask−viral infection RCT studies (Jefferson et al. 2020, Liu et al 2021): e.g., medical diagnosis of viral illness, self-reported symptoms of viral illness, lab-confirmed diagnosis of viral illness. Data from studies based on self-reported symptoms of viral illness were excluded because of awareness bias.

Awareness bias is tendency of a study participant to self-report a symptom or effect (e.g., a sickness or disease) because of concerns arising from prior knowledge of an environmental hazard that may cause the symptom (Schusterman 1992, Moffatt et al. 2000, Smith-Sivertsen et al. 2000, Rabinowitz et al. 2015). Participants in a study, where self-reporting is used to capture outcome measures, tend to overestimate their symptoms because of awareness bias.

Perception of exposure, causal beliefs and concerns, and media coverage have a role in study participants self-reporting symptoms (Borlee et al. 2019). Separating a true biological effect from reporting that is increased because of awareness bias is a problem in communities where study participants are aware of their potential exposure (Moffatt & Bhopal 2000).

Marcon et al. (2015) recommended using objective health outcomes to rule out awareness bias in populations potentially exposed to environmental hazards. Self-reported symptoms of viral illness cannot be considered objective unless it can be corroborated with other more credible outcome measures (i.e., laboratory confirmation) as such objectively measured outcomes are not influenced by awareness bias (Michaud et al. 2018).

Two online data bases – The Cochrane Central Register of Controlled Trials (CENTRAL) and PubMed – were used to identify eligible studies. These data bases were searched for meta-analysis or systematic reviews of randomized controlled trials investigating medical face mask use and Influenza or COVID (sars-cov-2) infections published from January 1, 2020 up to December 7, 2022.

The CENTRAL search strategy was relaxed in that it excluded targeted search terms such as mask, masks, facemasks, nonpharmaceutical, randomized or randomised. Here it was anticipated that there would not be many listings in the CENTRAL data base. The search was performed using the following terms: "influenza A" OR "influenza B" OR "covid" OR "sars-cov-2" OR "respiratory" in Title Abstract Keyword AND "infectious disease" Topic AND "01 January 2020 to 07 December 2022" Custom date range.

Due to potentially large number of COVID meta-analysis studies in the PubMed data base, the search strategy differed, and it included more targeted terms. These terms included: (((((((influenza[Title]) OR (covid[Title])) OR (sars-cov-2[Title])) OR (respiratory[Title])) OR (viral transmission[Title])) AND ((((nonpharmaceutical[Title]) OR (mask[Title])) OR (masks[Title])) OR (facemasks[Title]))) AND ((randomized[Title/Abstract]) OR (randomised[Title/Abstract]))) AND (("2020/01/01"[Date - Entry] : "2022/12/07"[Date - Entry])).

A potentially eligible systematic review was identified in gray literature during online searches. This review was published by the CATO Institute (Washington, DC) during the 01 January 2020 to 07 December 2022 period. This review was not captured by searches of the CENTRAL or PubMed data bases. It examined RCTs of medical mask use and viral (including Influenza and COVID) infections.

For each study identified through the searches, titles and full abstracts were read online. Based upon this, electronic copies of eligible meta-analysis or systematic review studies were then downloaded and read. The following criteria was used to determine eligibility of studies for the evaluation:

- Base studies were randomized controlled trials (RCTs) or cluster RCTs.
- Meta-analysis or systematic review.
- Compared the efficacy of medical masks with not wearing masks. Studies were excluded if they did not specify mask type used or present isolated outcomes for individual mask types.

- Included Influenza and/or COVID (sars-cov-2) viruses. Studies were excluded if they did not present isolated outcomes for these viruses.
- Intervention and control groups included community participants. Studies were excluded if they only involved workers in healthcare settings or they did not present isolated outcomes for community participants.
- Included credible outcome measures, i.e., medical diagnosis of viral illness or lab-confirmed diagnosis of viral illness.

**P-value plots**

In epidemiology it is traditional to use risk ratios or odds ratios and confidence intervals instead of p-values from a hypothesis test to demonstrate or interpret statistical significance. Both confidence intervals and p-values are constructed from the same data and they are inter-changeable. Altman & Bland (2011a,b) provide formulae showing how one can be calculated from the other. Standard statistical software packages – such as SAS and JMP (SAS Institute, Cary, NC) or STATA (StataCorp LLC, College Station, TX) – can also be used to estimate p-values from risk ratios or odds ratios and confidence intervals.

P-values were estimated using JMP statistical software from risk ratios or odds ratios and confidence intervals for all data in each of the studies evaluated. P-value plots after Schweder & Spjøtvoll (1982) were developed to inspect the distribution of the set of p-values for each study. The p-value is a random variable derived from a distribution of the test statistic used to analyze data and to test a null hypothesis (Young & Kindzierski, 2022).

In a well-designed and conducted study, the p-value is distributed uniformly over the interval 0 to 1 regardless of sample size under the null hypothesis (Schweder & Spjøtvoll, 1982). Suitably scaled, a distribution of p-values plotted against their ranks in a p-value plot should form a 45-degree line when there are no effects (Schweder & Spjøtvoll, 1982; Hung et al., 1997; Bordewijk et al., 2020). Researchers can use a p-value plot to inspect the heterogeneity of the test statistics combined in a meta-analysis.

The p-value plots constructed were interpreted as follows (Young & Kindzierski, 2022):
- Computed p-values were ordered from smallest to largest and plotted against the integers, 1, 2, 3,…
- If p-value points on the plot followed an approximate 45-degree line, it is concluded that test statistics resulted from a random (chance) process and the data supported the null hypothesis of no significant association or effect.
- If p-value points on the plot followed approximately a line with a flat/shallow slope, where most (the majority) of p-values were small (< 0.05), then test statistic data set provided evidence for a real, statistically significant, association or effect.
- If p-value points on the plot exhibited a bilinear shape (divided into two lines), the data set of test statistics used for meta-analysis is consistent with a two-component mixture and a general (overall) claim is not supported. In addition, a small p-value reported for the overall claim in the meta-analysis may not be valid (Schweder & Spjøtvoll, 1982).

Examples of p-value plots are provided in Appendix A after Young et al. (2022) to assist in interpretation of the p-value plots constructed here. Specifically, p-value plots in Appendix A represent 'plausible null' and 'plausible true alternative' hypothesis outcomes based on meta-analysis studies of observational data sets. As shown in the p-value plots in Appendix A:

- A plausible null hypothesis plots as an approximate 45-degree line.
- A plausible true alternative hypothesis plots as a line with a flat/shallow slope, where most (the majority) of p-values are small ($< 0.05$).

The distribution of the p-value under the alternative hypothesis – where p-values are a measure of evidence against the null hypothesis – is a function of both sample size and the true value or range of true values of the tested parameter (Hung et al., 1997). The p-value plots presented in Appendix A represent examples of distinct (single) sample distributions for each condition – i.e., for null (chance or random) associations and true effects between two variables. Evidence for p-value plots exhibiting behaviors outside of that shown in Appendix A should be treated as ambiguous (uncertain). A research claim based on ambiguous evidence is unproven.

**RESULTS**
**Search results**
CENTRAL – Sixty-one Cochrane Reviews published for the 01 January 2020 to 07 December 2022 period were identified. These search results are listed in Appendix B. From examining full abstracts for these reviews online, one eligible meta-analysis study that met the search criteria was found – Jefferson et al. (2021).

PubMed (medical research literature) – From the PubMed search, 73 records published for the period (refer to Appendix B) were identified. From examining full abstracts for these studies online, six eligible meta-analysis studies that met the search criteria were found – Aggarwal et al. (2020), Xiao et al. (2020), Nanda et al. (2021), Tran et al. (2021a), Kim et al. (2022), and Ollila et al (2022). Coincidentally, the Xiao et al. (2020) meta-analysis used the exact same RCT data as an earlier World Health Organization study (WHO 2019).

Gray literature – A final study included from gray literature was Liu et al. (2021), a systematic review by the public policy research organization CATO Institute.

**Eligible study attributes**
All eight studies evaluated are discussed, but only results for three of the eight studies are presented here – the Cochrane Review study (Jefferson et al. 2020), one study from medical research literature (Tran et al. 2021), and the study from gray literature (Liu et al. 2021). Results for the other five studies are presented in Appendix C.

*Cochrane review literature*
Jefferson et al. (2020) – Jefferson et al. ran computer searches in 6 data bases:
- Cochrane Central Register of Controlled Trials (CENTRAL) (2020, Issue 3)
- PubMed (2010 to 1 April 2020)
- The biomedical research data base Embase (2010 to 1 April 2020)
- CINAHL (Cumulative Index to Nursing and Allied Health Literature) (2010 to 1 April 2020)
- US National Institutes of Health Ongoing Trials Register ClinicalTrials.gov (January 2010 to 16 March 2020)

- World Health Organization International Clinical Trials Registry Platform (January 2010 to 16 March 2020)

Jefferson et al. identified and further analyzed 15 community (i.e., non-healthcare worker) RCTs – base studies – comparing medical masks versus no masks using the generalised inverse-variance random-effects model. The viral illness outcomes they reported were: numbers of acute respiratory infections, Influenza-like illness (ILI), laboratory-confirmed Influenza (LCI), or other viral pathogens. The specific focus of this evaluation was on data for numbers of ILI and LCI. This included nine ILI and six LCI outcomes (Analysis 1.1, p143, Jefferson et al. 2020). All of these data met the eligibility criteria.

Table 1 shows results for the 15 RCT base studies – primary outcome measures (risk ratio and 95% confidence intervals) and p-values that were estimated. Their research claim, i.e., cause−effect scientific claim, was (Authors' conclusions, p3, Jefferson et al. 2020)… "*pooled results of randomised trials did not show a clear reduction in respiratory viral infection with the use of medical/surgical masks during seasonal influenza*".

Table 1. Outcome measures (risk ratio and 95% confidence intervals) and p-values for 15 randomized control trials (base studies) included in Jefferson et al. (2020) meta-analysis.

| Outcome measure | 1st Author Year | Risk ratio (95% CI) | p-value |
|---|---|---|---|
| Influenza-like Illness (ILI) | Aiello 2012 | 1.10 (0.88 − 1.38) | 0.43304 |
| " | Barasheed 2014 | 0.58 (0.32 − 1.04) | 0.02222 |
| " | Canini 2010 | 1.03 (0.52 − 2.00) | 0.93667 |
| " | Cowling 2008 | 0.88 (0.34 − 2.27) | 0.80744 |
| " | Jacobs 2009 | 0.88 (0.02 − 31.84) | 0.98821 |
| " | MacIntyre 2009 | 1.11 (0.64 − 1.91) | 0.73421 |
| " | MacIntyre 2015 | 0.26 (0.03 − 2.51) | 0.24213 |
| " | MacIntyre 2016 | 0.32 (0.03 − 3.11) | 0.38679 |
| " | Suess 2012 | 0.61 (0.20 − 1.87) | 0.35996 |
| Lab-confirmed Influenza (LCI) | Aiello 2012 | 0.92 (0.59 − 1.42) | 0.70556 |
| " | Cowling 2008 | 1.16 (0.31 − 4.34) | 0.87632 |
| " | MacIntyre 2009 | 2.51 (0.74 − 8.50) | 0.44559 |
| " | MacIntyre 2015 | 0.83 (0.45 − 1.56) | 0.54827 |
| " | MacIntyre 2016 (1) | 0.97 (0.06 − 15.51) | 0.99393 |
| " | Suess 2012 | 0.39 (0.13 − 1.19) | 0.02408 |

*Medical research literature*
Aggarwal et al. (2020) – Aggarwal et al. (2020) ran computer searches on 25 April 2020 in two data bases: PubMed and Embase. They identified 902 records from their searches. They undertook reviews of 83 full text articles, of which 74 were excluded by their criteria. The remaining nine studies (cluster-RCTs) were used by Aggarwal et al. for their meta-analysis. Five of these studies were for medical mask versus no mask use by community participants. Their meta-analysis used the random effects model. Viral illness outcomes they reported were ILI, self-reported ILI, and LCI.

Aggarwal et al. results for the 5 cluster-RCT base studies are shown elsewhere (Table C1, Appendix C). These include outcome measures (effect sizes and 95% confidence intervals) and p-values that we estimated. Two of the five outcome measures failed to meet the eligibility criteria as they were based on self-reported ILI (with attendant awareness bias) (Table C1). Consequently, Aggarwal et al. results were not used for p-value plotting. Their research claim was (Abstract, Aggarwal et al. 2020)… "*data pooled from randomized controlled trials do not reveal a reduction in occurrence of ILI with use of facemask alone in community* settings".

Xiao et al. (2020) – Xiao et al. investigated multiple nonpharmaceutical measures (hand hygiene, masks) for pandemic influenza in nonhealthcare (community) settings. For the 'mask' component of their investigation, they ran computer searches in four databases (CENTRAL, PubMed, Embase, Medline) to identify 'randomized controlled trial in community setting' studies that were available from 1946 through July 28, 2018. They identified and screened the titles of 1,100 articles, from which 856 were excluded.

Xiao et al. reviewed full abstracts of the remaining 244 articles and undertook reviews of 98 full text articles. From this list, 10 RCT articles were identified in which seven RCTs were included as base studies in meta-analysis of medical mask versus no mask using the fixed effects model. The viral illness outcome they reported was LCI.

Xiao et al. results for the seven RCT base studies are shown elsewhere (Table C2, Appendix C). These include– outcome measure (risk ratio and 95% confidence intervals) and p-values that were estimated. Their research claim was (Abstract, Xiao et al. 2020)… "*Although mechanistic studies support the potential effect of hand hygiene or face masks, evidence from 14 randomized controlled trials of these measures did not support a substantial effect on transmission of laboratory-confirmed influenza*".

Incidentally, the Xiao et al. meta-analysis and results replicates exactly an earlier World Health Organization (WHO 2019) investigation of mask use related to epidemic and pandemic influenza. WHO (2019) used the exact same seven base studies in a meta-analysis and reported the exact same quantitative results. The WHO research claim was (Executive Summary, WHO 2019)… "*There are a number of high-quality randomized controlled trials demonstrating that personal measures (e.g. hand hygiene and face masks) have at best a small effect on transmission*".

Nanda et al. (2021) – Nanda et al. evaluated RCTs of cloth and medical face mask use (± hand hygiene) for preventing respiratory virus transmission in the community setting. They ran computer searches in three databases (CENTRAL, PubMed, Embase). They identified and screened the titles of 1,499 articles, from which 1,126 were excluded. They reviewed full texts of 373 articles. From this list, 11 RCT articles were included as base studies in their meta-analysis. The viral illness outcome they reported was laboratory-confirmed virus.

Also incidentally, for meta-analysis of RCTs comparing masks alone to no masks for laboratory confirmed virus, Nanda et al. identified and used the exact same seven base studies as WHO (2019) and Xiao et al. (2020). However, data extracted by Nando et al. from the base studies and used for calculating risk ratios and confidence interval differed compared to WHO (2019) and Xiao et al. (2020).

Nanda et al. results for the seven RCT base studies are shown elsewhere (Table C3, Appendix C). These include outcome measure (risk ratio and 95% confidence intervals) and p-values that were estimated. Their research claim was (Abstract, Nando et al. 2021)… "*There is limited available preclinical and clinical evidence for face mask benefit in sars-cov-2. RCT evidence for other respiratory viral illnesses shows no significant benefit of masks in limiting transmission*"

<u>Tran et al. (2021)</u> – Tran et al. registered a protocol for their study in PROSPERO on 7 May 2020 (Tran et al. 2020). They performed a systematic review and network meta-analysis of RCTs to assess efficacy of face masks in preventing respiratory infections in community settings. They ran computer searches in nine databases – CENTRAL, PubMed, Embase, Web of Science (ISI), Scopus, Google Scholar, ASSIA, Clinicaltrials.gov, System for Information on Grey Literature in Europe (SIGLE).

They identified and screened the titles and abstract of a total of 13,988 articles, from which 13,876 were excluded. They reviewed full texts of 112 articles, plus they added 1 article from gray literature. From this list, 16 RCT articles were selected for their overall analysis and eight RCT articles were included as base studies in their mask versus no mask meta-analysis. Tran et al. used the fixed effects model in their meta-analysis. The viral illness outcome they reported was ILI. Seven of the eight RCT base studies used in their meta-analysis were the exact same as those used by WHO (2019), Xiao et al. (2020), and Nanda et al. (2021). The one additional base study they used was Canini et al. (2010).

Tran et al. results for the eight RCT base studies are shown in Table 2. This includes outcome measure (risk ratio and 95% confidence intervals) and p-values that were estimated. Their research claim was (Abstract, Tran et al., 2021)… "*Given the body of evidence through a systematic review and meta-analyses, our findings supported the protective benefits of MFMs* [medical face masks] *in reducing respiratory transmissions, and the universal mask-wearing should be applied—especially during the COVID-19 pandemic*".

Table 2. Outcome measures (risk ratio and 95% confidence intervals) and p-values for 8 randomized control trials (base studies) included in Tran et al. (2021) meta-analysis.

| Outcome measure | 1st Author Year | Risk ratio (95% CI) | p-value |
|---|---|---|---|
| Influenza-like Illness (ILI) | Aiello 2010 | $0.78 \ (0.64 - 0.96)$ | 0.007 |
| " | Aiello 2012 | $0.85 \ (0.58 - 1.24)$ | 0.373 |
| " | Barasheed 2014 | $0.58 \ (0.33 - 1.01)$ | 0.0155 |
| " | Canini 2010 | $1.02 \ (0.61 - 1.71)$ | 0.9432 |
| " | Cowling 2008 | $2.05 \ (0.69 - 6.04)$ | 0.4417 |
| " | MacIntyre 2009 | $1.31 \ (0.72 - 2.40)$ | 0.4695 |
| " | MacIntyre 2016 | $0.33 \ (0.03 - 3.11)$ | 0.0116 |
| " | Suess 2012 | $0.51 \ (0.21 - 1.25)$ | 0.0648 |

<u>Kim et al. (2022)</u> – Kim et al. initially registered a protocol for their study in PROSPERO on 28 October 2020 and changed the protocol on 20 November 2020 (Seong et al. 2020). They performed a network meta-analysis of RCTs to assess efficacy of face masks in preventing respiratory infections in community settings. They ran computer searches in PubMed, Google Scholar and medRxiv data bases for studies published up to 5 February 2021.

They identified and screened the titles of 5,946 articles, from which 5,761 were excluded. They reviewed full texts of 185 articles. From this list, 35 articles were selected for their overall analysis; which included RCTs, prospective cohort studies, retrospective cohort studies, case–control studies and cross-sectional studies.

The focus here is on RCTs used as base studies in their mask versus no mask meta-analysis. Kim et al. used the random effects model in their meta-analysis. The viral illness outcome they reported was LCI for Influenza (6 base studies) and lab-confirmed infection for COVID (1 base study).

Kim et al. results for the seven RCT base studies are shown elsewhere (Table C4, Appendix C) shows). These include outcome measure (risk ratio and 95% confidence intervals) and p-values that were estimated. Their research claim was (Abstract, Kim et al. 2022)… "*Evidence supporting the use of medical or surgical masks against influenza or coronavirus infections (SARS, MERS and COVID-19) was weak*".

Ollila et al. (2022) – Ollila et al. initially registered a protocol for their study in PROSPERO on 16 November 2020 and changed the protocol on 12 May 2022 and again on 22 September 2022 (Ollila et al. 2020) before it was published on 1 December 2022. They performed a systematic review and meta-analysis of RCTs to assess efficacy of face masks in preventing respiratory infections in community settings.

They ran computer searches in CENTRAL, PubMed, Embase, and the Web of Science data bases for studies published between 1981 and 9 February 2022. It is noted that well into their study (initially registered 16 November 2020), they first changed the research protocol 16 months later (12 May 2022) and then again four months later.

They identified and screened 1,836 articles, from which 1,785 were excluded. They reviewed full texts of 49 articles. From this list, 18 RCT articles were selected for their analysis; eight of these were specific to community settings and 10 were specific to non-community settings. Here, interest was in the eight results for community settings.

Ollila et al. results for the eight RCT base studies are shown elsewhere (Table C5, Appendix C). This just includes outcome measures (odds ratio and 95% confidence intervals), not p-values. Their research claim was (Abstract, Ollila et al. 2022)… "*Our findings support the use of face masks particularly in a community setting and for adults*".

P-values were not estimated for base study statistics used by Ollila et al. Six of the eight outcome measures failed to meet the eligibility criteria. Specifically, five of these measures were based on self-reported symptoms (with attendant awareness bias), and the origin of one measure Ollila et al. used for another base study could not be confirmed (Table C5). This was determined by accessing and reading each of the eight base studies used by Ollila et al.

Based on our readings of the base studies and the fact that they changed their protocol twice well into the study before it was published, there are concerns about the reliability of this meta-analysis. Nowhere in the meta-analysis do they state which outcome measures were used.

These practices – changing the research protocol multiple times and failing to indicate specific outcome measures in their paper – imply selective analysis and reporting. Researchers have flexibility to use different (but valid) methods in a study. Unfortunately, they then have further flexibility to only report those methods that yield favorable results and ignore those that yield unfavorable results (Kavvoura et al. 2007, Ioannidis 2008, Contopoulos-Ioannidis et al. 2009, Ioannidis et al. 2011, Carp 2012).

This reporting preference involves selective tendency to highlight statistically significant findings and to avoid highlighting nonsignificant findings in research (Kavvoura et al. 2007). This can be problematic because the significant findings could in future turn out to be false positives.

Also, test statistics used for three of the base studies for self-reported symptoms showing a benefit of mask use – Barasheed et al. (2014), Aiello et al. (2010a) and Abaluck et al. (2021) in Table C5 – are opposite to other published data of lab-confirmed statistics for the same studies.

Specifically, for each of these base studies Ollila et al. reported a significant difference between mask and control group outcomes in their meta-analysis; whereas published data exist for laboratory-confirmed infections showing no difference between the mask and control groups. For Barasheed et al. (2014), data reported for lab-confirmed infections in their study showed no difference. For Aiello et al. (2010a), Aiello et al. (2010b) reported that lab-confirmed infections for the same study showed no difference.

For Abaluck et al. (2021), Chikina et al. (2022) independently reviewed this study and identified numerous biases unreported by Abaluck et al. that complicate inferences of causality. Chikina et al.'s own analysis of the study data identified a difference of just 20 lab-confirmed COVID cases between the mask and no mask groups in a study population of over 300,000 individuals (i.e., 1,106 COVID symptomatic seropositives in the mask group versus 1,086 in the no mask group).

Chikina et al. stated that… "*it would not be reasonable to conclude from this trial that there is a direct causal link between mask wearing and the number of residents in villages and households, any causal claims based on effects of similar size in this trial should be considered with caution*".

A final observation is the "main" result reported by Abaluck et al. (2021) themselves in their study (Results, page 1)… "*Adjusting for baseline covariates, the intervention* [masking] *reduced symptomatic seroprevalence by 9.5% (adjusted prevalence ratio = 0.91 [0.82, 1.00]*". This result is not significant (p-value=0.062).

As the self-reported statistics used by Ollila et al. for the Abaluck et al. (2021) study was opposite to more-reliable lab-confirmed statistics, and given other biases, the study itself and their claim … *"support the use of face masks particularly in a community setting and for adults"* is judged unreliable.

*Gray literature*

Liu et al. (2021) – The Liu et al. systematic review involved examining available clinical evidence of the impact of face mask use in community settings on respiratory infection rates, including by COVID. This review was different than other meta-analysis evaluated here in that it did not specify methodologies for identification of RCT base studies. However, they did present and discuss the results of RCTs that they identified.

As a result of their different methodology, an attempt was made to obtain original copies of the base studies to confirm results reported by Liu et al. They reported outcome measures as p-values for 16 RCT base papers. Only 14 of the 16 base papers were obtained. Lui et al. results for the 14 base papers are presented elsewhere (Table C6, Appendix C) because additional supporting information provided made the table too big to present here.

With respect to their results, specific interest was on data for clinical diagnosis of ILI, and LCI or laboratory-confirmed other viral pathogens. Multiple test statistics were identified in the 14 base papers. These statistics were converted to p-values and presented in Table C6 (p-values used for plotting are highlighted, bolded and italicized).

Their research claim was (Abstract, Liu et al. 2022)… *"Of sixteen quantitative meta-analyses, eight were equivocal or critical as to whether evidence supports a public recommendation of masks, and the remaining eight supported a public mask intervention on limited evidence primarily on the basis of the precautionary principle"*.

**P-value plots**

Six p-value plots were constructed, three of which are presented here: the Jefferson et al. (2020) (Figure 2) and Tran et al. (2021) (Figure 3) meta-analyses, and the Liu et al. (2021) systematic review (Figure 4). Three other p-value plots are presented elsewhere (Appendix C) for the Xiao et al. (2020) (Figure C1), Nanda et al. (2021) (Figure C2), and Kim et al. (2022) (Figure C3) meta-analyses. P-value plots for Aggarwal et al. (2020) and Ollila et al. (2022) meta-analyses were not constructed because of overreliance on self-reported outcomes in their meta-analyses or irregularities (biases) discussed above.

For all the plots presented here and in Appendix C, no evidence was observed of distinct (single) sample distributions for true effects between two variables (i.e., p-value points forming a line with a flat/shallow slope, where most (the majority) of p-values are small, < 0.05). Again, the reader is referred to Appendix A for examples of p-value plots showing true effects between two variables in observational studies.

The Jefferson et al. (2020) (Figure 2) and Liu et al. (2021) (Figure 4) meta-analyses show evidence of distinct (single) sample distributions for null effects – chance or random associations – between two variables (i.e., p-value points plot as an approximate 45-degree line).

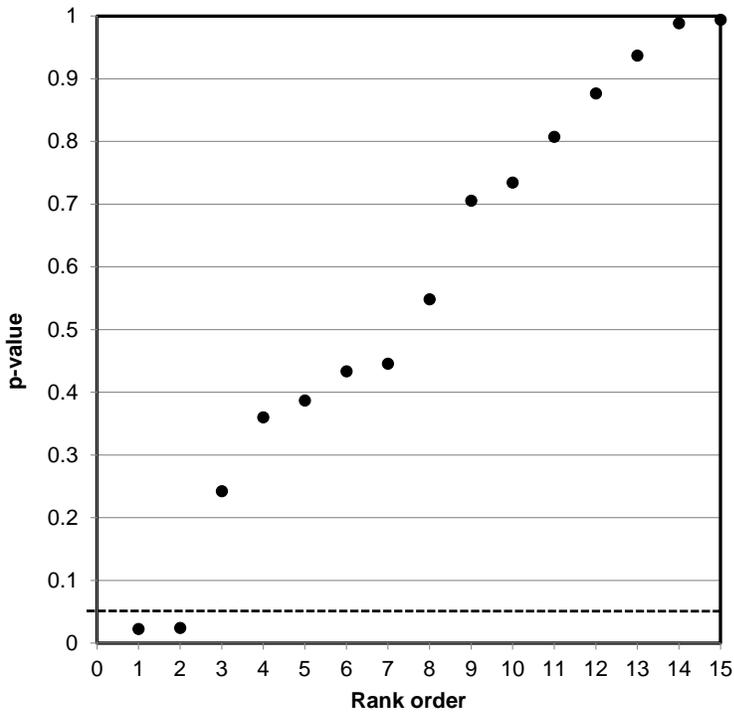

Figure 2. p-value plot for 15 randomized control trials (base studies) included in the Jefferson et al. (2020), Cochrane review, meta-analysis.

The Trans et al. (2021) meta-analysis (Figure 3) exhibits a bilinear shape (divides into two lines) – three p-values are small (<0.05) and five p-values >0.05 are oriented on an approximate 45-degree line. This data set of test statistics is consistent with a two-component mixture and thus a general (overall) claim is unproven.

The Xiao et al. (2020) (Figure C1) and Kim et al. (2021) (Figure C3) p-value plots are only based on seven points and yet both show evidence of distinct (single) sample distributions for null effects – chance or random associations – between two variables. The Nanda et al. (2021) p-value points (Figure C2) plot closer to a 40-degree line. However, it is still clearly supportive of null effects versus true effects.

P-values are interchangeable with traditional epidemiology risk statistics (i.e., risk ratios or odds ratios and confidence intervals). Table 3 presents a summary of p-values that were estimated for risk statistics drawn from base studies used in six meta-analyses. P-values were not estimated for the Ollila et al. (2022) meta-analysis and p-values for the Liu et al. (2021) systematic review are not shown. Including those listed in Table 3, Table C5 (Ollila et al. 2022) and Table C6 (Liu et al. 2021), a total of 18 base studies were used across the seven meta-analyses and one systematic review.

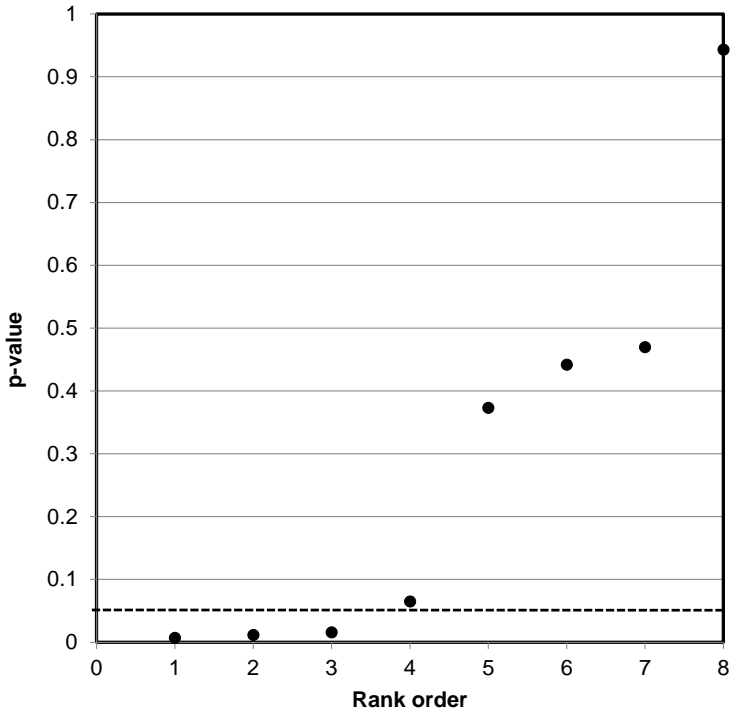

Figure 3. p-value plot for 8 randomized control trials (base studies) included in the Tran et al. (2020) meta-analysis.

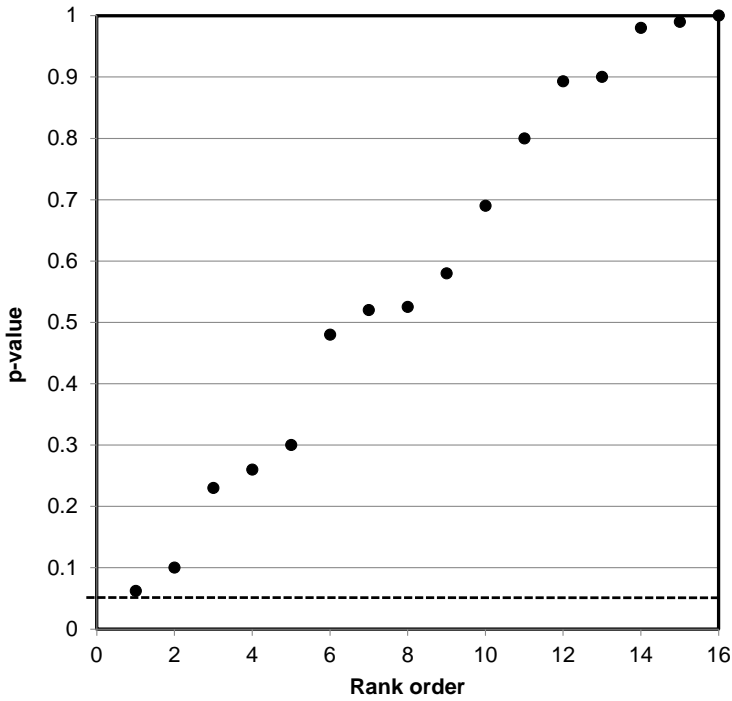

Figure 4. p-value plot for 16 randomized control trials (base studies) included in the Liu et al. (2021), systematic review, meta-analysis.

Table 3. Summary of p-values used in six meta-analysis studies.

| Meta-analysis:<br>Base study,<br>1st Author Year | Jefferson et al. | Aggarwal et al. | Xiao et al. | Nanda et al. | Tran et al. | Kim et al. |
|---|---|---|---|---|---|---|
| Aiello 2010 | | 0.0369 | 0.5663 | 0.1926 | 0.007 | |
| Aiello 2012 | 0.4334, 0.7056 | 0.5046 | 0.3187 | 0.4368 | 0.373 | 0.3148 |
| Alfelali 2020 | | | | | | 0.7452 |
| Barasheed 2014 | 0.0222 | | 0.8815 | 0.2095 | 0.0155 | |
| Bundgaard 2020 | | | | | | 0.2994 |
| Canini 2010 | 0.9367 | | | | 0.9432 | |
| Cowling 2008 | 0.8074, 0.8763 | 0.2812 | 0.8746 | 0.8063 | 0.4417 | 0.8763 |
| Jacobs 2009 | 0.9882 | | | | | |
| MacIntyre 2009 | 0.7342, 0.4456 | 0.4744 | 0.9115 | 0.3404 | 0.4695 | 0.8671 |
| MacIntyre 2015 | 0.2421, 0.5483 | | | | | |
| MacIntyre 2016 | 0.3868, 0.9939 | | 0.7411 | 0.485 | 0.0116 | |
| Simmerman 2011 | | | | | | |
| Suess 2012 | 0.36, 0.0241 | 0.4785 | 0.0009 | 0.0167 | 0.0648 | 0.0002 |

Recall that a meta-analysis first involves a systematic review. The meta-analysis then integrates results of identified studies from the systematic review. One would anticipate that well-conducted, independent meta-analysis studies examining the same research question – does medical mask use in community settings prevent COVID infection– should identify similar or even the same base studies published within the same period for their analysis. Table 3 shows that while most of the base studies used are similar across the meta-analyses, they are not the same.

An inconsistency apparent in Table 3 is that different data is being drawn by various independent meta-analysis researchers from a base study for the exact same research question. Take the Aiello 2010 base study, which is used in four meta-analyses (Table 3). Two meta-analyses used risk statistics that are significant (i.e., p-value < 0.05) – Aggarwal et al. and Tran et al. The other two meta-analyses used risk statistics that are non-significant (i.e., p-value > 0.05) – Xiao et al. and Nanda et al.

This raises a question of why different quantitative results are used by meta-analysis researchers examining the same research question? Is it due to selective analysis and reporting of the researchers or some other limitation of the meta-analysis process itself?

## DISCUSSION AND IMPLICATIONS

An objective of this evaluation was to evaluate reproducibility of research claims in meta-analysis or systematic review studies of mask use in community settings to prevent COVID infection. Eight eligible studies – seven meta-analyses and one systematic review – were identified and evaluated.

These studies were published between the period 1 January 2020 to 7 December 2022. P-value plots were constructed to visually inspect the heterogeneity of test statistics combined in these studies. Table 4 compares research claims made in the seven meta-analysis and one systematic review to findings using p-value plots.

A true effect between two variables in meta-analysis should comprise a set of homogeneous (similar) statistics that represent a distinct (single) sample distribution in a p-value plot. This type of effect should show points that align with a shallow slope in the plot. A null effect (chance or random association) between two variables in meta-analysis should show points uniformly distributed over the interval 0 to 1 regardless of sample size in a p-value plot. This type of effect should show points aligned approximately 45 degrees in the plot.

Table 4. Comparison of meta-analysis research claims to independent results using p-value plots.

| Study detail* | Study research claim[+] | Independent finding of p-value plot | Is study research claim supported? |
|---|---|---|---|
| Cochrane review literature: | | | |
| Jefferson et al. (2020) meta analysis | no significant benefit to medical mask use | null (no) effect | yes |
| Medical research literature: | | | |
| Aggarwal et al. (2020) meta-analysis | no significant benefit to medical mask use | insufficient data to examine | unable to determine |
| Xiao et al. (2020) meta-analysis | " | null effect | yes |
| Nanda et al. (2021) meta-analysis | " | null effect | yes |
| Tran et al. (2021) meta-analysis | benefit to medical mask use | finding is ambiguous (uncertain) | no |
| Kim et al. (2022) meta-analysis | " | null effect | no |
| Ollila et al. (2022) meta-analysis | " | insufficient data to examine | unable to determine |
| Grey literature: | | | |
| Liu et al. (2021) systematic review | no significant benefit to medical mask use | null effect | yes |

**\*** All studies examined randomized control trials of medical mask versus no mask use in community settings for reduction of viral infection (Influenza or COVID virus).

[+] benefit ≡ reduces viral infection.

For six p-value plots constructed (five meta-analyses and one systematic review), no evidence of distinct sample distributions for true effects between two variables was observed. Five of these plots showed points aligned approximately with 45 degrees – indicating null effects. These p-value plots are consistent with chance or random associations (i.e., no proven benefit) for medical mask use in community settings to prevent viral, including COVID, infection.

One other plot (data set of Tran et al. 2021, Figure 3) had p-value points divided into two lines; consistent with a heterogenic or dissimilar data set (two-component mixture). Here there is insufficient evidence to make a research claim because of ambiguity (uncertainty) in the data set used for meta-analysis.

P-value plots for two other meta-analyses – Aggarwal et al. (2020) and Ollila et al. (2022) – were not constructed because of overreliance on self-reported outcomes (with attendant awareness bias) and other irregularities (i.e., biases).

Wang et al. (2021) present ample evidence of airborne transmission for many respiratory viruses. These include Influenza virus, respiratory syncytial virus (RSV), human Rhinovirus, severe acute respiratory syndrome coronavirus (SARS-CoV), Middle East respiratory syndrome coronavirus (MERS-CoV), SARS-CoV-2 (COVID), measles virus, adenovirus, and enterovirus.

COVID RNA fragments have been identified and infectious COVID virus has been found in airborne aerosols from 0.25 to >4 mm (Wang et al. 2021). This is consistent with that observed for the Influenza virus, where RNA has been identified in both ≤5 μm and >5 μm aerosols respired from infected hosts, with more Influenza virus RNA found in the ≤5 μm aerosols (Fennelly 2020, Wang et al. 2021). The World Health Organization chief scientist recently acknowledged that the COVID was an airborne virus spread by aerosols (Kupferschmidt 2022).

These observations highlight the importance of airborne aerosol transmission and infection for respiratory viruses, including COVID. Medical mask RCTs of Influenza infection are directly applicable for understanding the benefit of their use to prevent COVID infection. Again, it is not the virus itself but airborne transmission of aerosols or droplets containing viruses that is important for infection.

Where observational data are used in randomized (or even non-randomized) medical intervention studies, a strong statistical component is required to establish informative and interpretable intervention−risk/benefit associations. This is also the case for research claims made from these associations. For a research claim to be considered valid, it must defeat randomness (i.e., a statistical outcome due to chance).

The p-value plots for five studies – Jefferson et al. (2020), Figure 2; Xiao et al. (2020), Figure C1; Nanda et al. (2021), Figure C2; Kim et al. (2022), Figure C3; Liu et al. (2021), Figure 4 – show results that look random. The findings of randomness are consistent with research claims made by Jefferson et al. (2020), Xiao et al. (2020), Nanda et al. (2021), and Lui et al. (2021), i.e., no significant benefit to medical mask use (refer to Table 4).

In short, p-value plots were able to reproduce and support their research claims. These reproducible results strengthen a claim that medical masks have unproven benefit in community settings to prevent respiratory virus infections. This has been reported in the past (Inglesby et al. 2005, Hardie 2016) and more recently (Drummond 2022, Miller 2022).

The p-value plot finding of randomness in Figure C3 is opposite to the research claim of Kim et al. (2022) (benefit to medical mask use). This implies irreproducibility of their claim. The reproducibility of claims by Aggarwal et al. (2020) (no benefit to medical mask use) and Ollila et al. (2022) (benefit to medical mask use) were not evaluated because of insufficient data for p-value plots. The latter meta-analysis is judged unreliable due to overreliance on self-reported outcomes (with attendant awareness bias) and irregularities (i.e., biases) discussed previously.

For an intervention to be useful and practical to a population, any benefit of the intervention must be of sufficient magnitude to be able to observe a difference in an outcome between the intervention group and a control group at the population level. Consider the countries of Germany and Sweden during the pandemic. Germany had a mask mandate for its population whereas Sweden did not.

Survey data on mask compliance during the pandemic was captured in many countries by the University of Maryland (UMD) Social Data Science Center working in collaboration with Facebook (UMD 2022). One of the survey questions asked Facebook users if they wore a mask most or all the time in the previous 5 days. Figure 5 shows Facebook user reported monthly average mask compliance (%) during the second COVID wave – October 2020 through June 2021 – in Germany and Sweden. Figure 5 shows that mask compliance in Germany was never less than 80%. Whereas mask compliance in Sweden was never more than 21%.

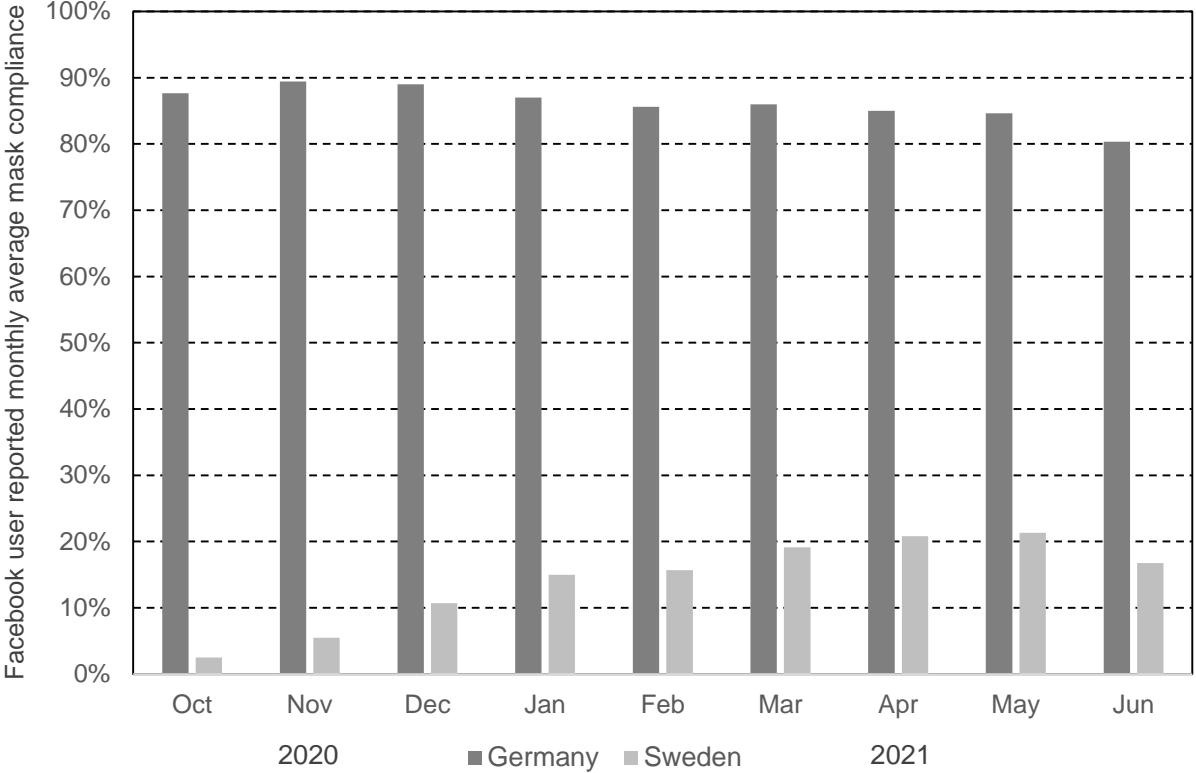

Figure 5. Facebook user reported monthly average mask compliance (%) during the second COVID wave in the countries of Germany and Sweden.

Note: Mask compliance data shown here is averaged from daily data representing percent of Facebook respondents that reported wearing a mask most or all the time in the previous 5 days; data are from the UMD COVID Trends and Impacts Survey (https://gisumd.github.io/COVID-19-API-Documentation/); data are adjusted by Facebook for selection biases (non-response and sampling frame coverage bias).

What role might have masks played during the second wave? Consider Figure 6, depicting daily new COVID deaths per million population in Germany and Sweden. Figure 6 was originally derived by Miller (2022) and is reproduced here using data from the World Health Organization COVID dashboard (https://covid19.who.int/).

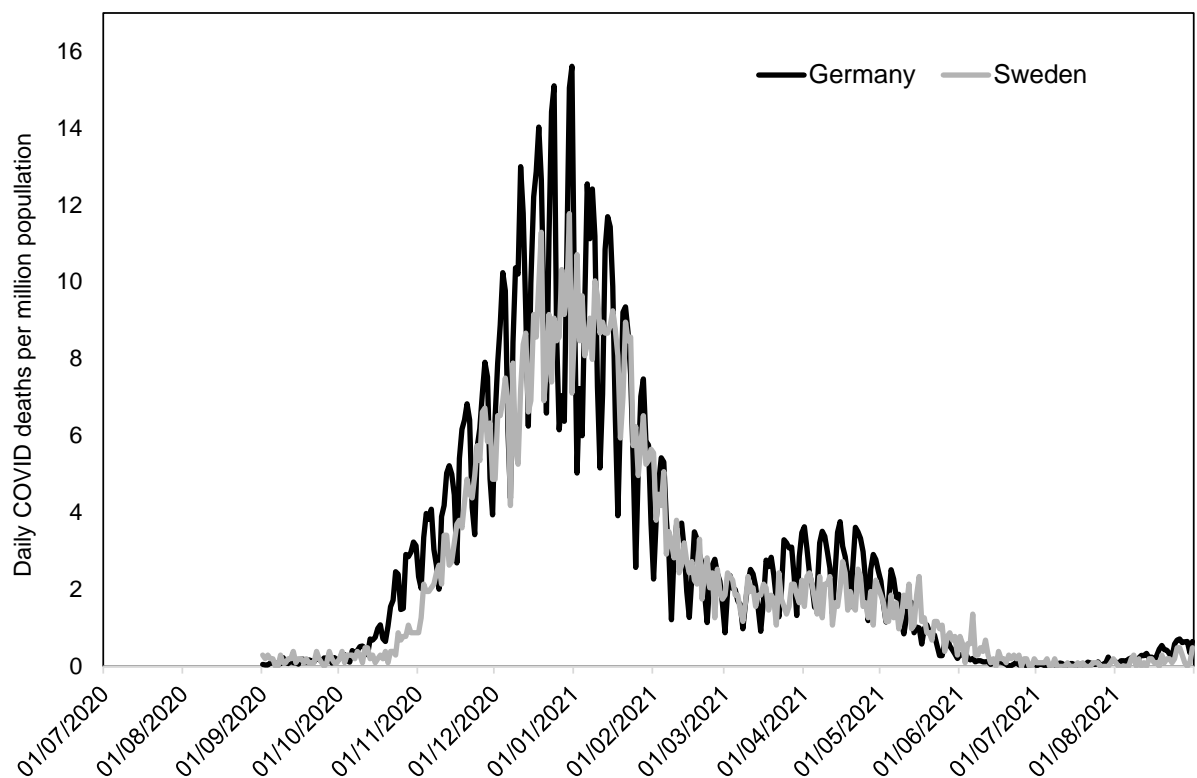

Figure 6. Daily new daily COVID deaths per million population during the second wave in Germany and Sweden.

   Note: Daily death data are from the WHO COVID dashboard (https://covid19.who.int/).

Figure 6 shows that WHO-reported daily COVID deaths per million population are not much different in the two countries. At the population level, a first impression of these two figures is that mask use had little or no benefit in preventing COVID deaths during the second wave. Despite obvious differences in mask compliance among these countries (Figure 5), reported daily COVID deaths per million population are not much different (Figure 6).

**ACKNOWLEDGEMENTS**

No external funding was provided for this study. The study was conceived based on previous work undertaken by CG Stat for the National Association of Scholars (nas.org), New York, NY.

**ORCHID ID**

S. Stanley Young: https://orcid.org/0000-0001-9449-5478
Warren B. Kindzierski: https://orcid.org/0000-0002-3711-009X

**REFERENCES**

Aggarwal, N., Dwarakanathan, V., Gautam, N., et al. (2020). Facemasks for prevention of viral respiratory infections in community settings: A systematic review and meta-analysis. Indian Journal of Public Health, 64(Supplement), S192–S200. https://doi.org/10.4103/ijph.IJPH_470_20

Altman, D. G., & Bland, J. M. (2011a). How to obtain a confidence interval from a P value. British Medical Journal, 343, d2090. https://doi.org/10.1136/bmj.d2090

Altman, D. G., & Bland, J. M. (2011b). How to obtain the P value from a confidence interval. British Medical Journal, 343, d2304. https://doi.org/10.1136/bmj.d2304

Bałazy, A., Toivola, M., Adhikari, A., et al. (2006). Do N95 respirators provide 95% protection level against airborne viruses, and how adequate are surgical masks? American Journal of Infection Control, 34(2), 51−57. https://doi.org/10.1016/j.ajic.2005.08.018

Bar-On, Y. M., Flamholz, A., Phillips, R., et al. (2020). SARS-CoV-2 (COVID-19) by the numbers. Elife, 9, e57309. https://doi.org/10.7554/eLife.57309

Begley, C. G., & Ioannidis, J. P. (2015). Reproducibility in science: Improving the standard for basic and preclinical research. Circulation Research, 116(1), 116−126. https://doi.org/10.1161/CIRCRESAHA.114.303819

Belkin, N. (1996). A century after their introduction, are surgical masks necessary? American Operating Room Nurses Journal, 64(4), 602−607. http://doi.org/10.1016/s0001-2092(06)63628-4

Bordewijk, E. M., Wang, R., Aski, L. M., et al. (2020). Data integrity of 35 randomised controlled trials in women' health. European Journal of Obstetrics & Gynecology and Reproductive Biology, 249, 72−83. http://dx.doi.org/10.1016/j.ejogrb.2020.04.016

Boos, D.D., & Stefanski, L.A. (2013). Essential Statistical Inference: Theory and Methods. New York, NY: Springer.

Borlee, F., Yzermans, C. J., Oostwegel, F. S. M., et al. (2019). Attitude toward livestock farming does not influence the earlier observed association between proximity to goat farms and self-reported pneumonia. Environmental Epidemiology, 3(2), e041. http://doi.org/10.1097/EE9.0000000000000041

Bramstedt, K. A. (2020). The carnage of substandard research during the COVID-19 pandemic: A call for quality. Journal of Medical Ethics, 46, 803–807. http://doi.org/10.1136/medethics-2020-106494.

Canini, L., Andreoletti, L., Ferrari, P., et al. (2010). Surgical mask to prevent influenza transmission in households: a cluster randomized trial. PLoS One, 5(11), e13998. http://doi.org/10.1371/journal.pone.0013998

Carp, J. (2012). The secret lives of experiments: Methods reporting in the fMRI literature. Neuroimage, 63(1), 289−300. http://dx.doi.org/10.1016/j.neuroimage.2012.07.004

Centers for Disease Control and Prevention (CDC). (2007). Interim pre-pandemic planning guidance: community strategy for pandemic influenza mitigation in the United States: Early, targeted, layered use of nonpharmaceutical interventions. Atlanta, GA: US CDC. https://stacks.cdc.gov/view/cdc/11425

Centers for Disease Control and Prevention (CDC). (2017). Community mitigation guidelines to prevent pandemic influenza – United States. Atlanta, GA: US CDC. MMWR Recommendations and reports, 66(No. RR-1)1−36. https://stacks.cdc.gov/view/cdc/45220

Centers for Disease Control and Prevention (CDC). (2022). CDC Museum COVID-19 Timeline, August 16, 2022. https://www.cdc.gov/museum/timeline/covid19.html

Chu, C., Baxamusa, S., & Witherel, C. (2021). Impact of COVID-19 on materials science research innovation and related pandemic response. MRS Bulletin, 46, 807–812. https://doi.org/10.1557/s43577-021-00186-1

Clase, C. M., Fu, E. L., Joseph, M., et al. (2020). Cloth masks may prevent transmission of COVID-19: An evidence-based, risk-based approach. Annals of Internal Medicine, 173(6), 489–491. https://doi.org/10.7326/M20-2567

Contopoulos-Ioannidis, D. G., Karvouni, A., Kouri, I., et al. (2009). Reporting and interpretation of SF-36 outcomes in randomised trials: Systematic review. British Medical Journal, 338, a3006. https://doi.org/10.1136/bmj.a3006

Drummond, H. (2022). The Face Mask Cult. ISBN: 978-1-9999907-9-4. UK: CantusHead Books. https://hectordrummond.com/the-face-mask-cult/

Egger, M., Davey Smith, G., & Altman, D. G. (2001). Problems and limitations in conducting systematic reviews. In: Egger, M., Davey Smith, G., & Altman, D. G. (eds.) Systematic Reviews in Health Care: Meta−analysis in Context, 2nd ed. London: BMJ Books.

Fennelly, K. P. (2020). Particle sizes of infectious aerosols: Implications for infection control. Lancet Respiratory Medicine, 8(9), 914−924. https://doi.org/10.1016/S2213-2600(20)30323-4

Furukawa, N. W., Brooks, J. T., & Sobel, J. (2020). Evidence supporting transmission of severe acute respiratory syndrome coronavirus 2 while presymptomatic or asymptomatic. Emerging Infectious Diseases, 26(7). https://doi.org/10.3201/eid2607.201595

Gostin, L. O., Friedman, E. A., & Wetter, S. A. (2020). Responding to covid-19: How to navigate a public health emergency legally and ethically. Hastings Center Report, 50, 8−12. http://doi.org/10.1002/hast.1090

Gustot, T. (2020). Quality and reproducibility during the COVID-19 pandemic. JHEP Reports, 2, 1−3. https://doi.org/10.1016/j.jhepr.2020.100141

Han, Z. Y., Weng, W. G., & Huang, Q. Y. (2013). Characterizations of particle size distribution of the droplets exhaled by sneeze. Journal of the Royal Society Interface, 10(88), 20130560. http://dx.doi.org/10.1098/rsif.2013.0560

Hardie, J. (2016). Why face masks don't work: A revealing review. Oral Health, October 18, 2018. https://web.archive.org/web/20200509053953/https:/www.oralhealthgroup.com/features/face-masks-dont-work-revealing-review/

Herner, M. (2019). Perfect top of the evidence hierarchy pyramid, maybe not so perfect: lessons learned by a novice researcher engaging in a meta-analysis project. BMJ Evidence-Based Medicine, 24(4), 130−132. https://doi.org/10.1136/bmjebm-2018-111141

Hung, H. M. J., O'Neill, R. T., Bauer, P., et al. (1997). The behavior of the p-value when the alternative hypothesis is true. Biometrics, 53, 11–22. https://doi.org/10.2307/2533093

Inglesby, T. V., Nuzzo, J. B. O'Tool, T., et al. (2006). Disease mitigation measures in the control of pandemic influenza. Biosecurity and Bioterrorism: Biodefense Strategy, Practice, and Science, 4(4), 366–375. https://doi.org/10.1089/bsp.2006.4.366

Ioannidis, J. P. A. (2005). Why most published research findings are false. PLoS Medicine, 2(8), e124. https://doi.org/10.1371/journal.pmed.0020124

Ioannidis J. P. (2008). Why most discovered true associations are inflated. Epidemiology, 19(5), 640–648. https://doi.org/10.1097/EDE.0b013e31818131e7

Ioannidis, J. P. A. (2022). Correction: Why most published research findings are false. PLoS Medicine, 19(8), e1004085. https://doi.org/10.1371/journal.pmed.1004085

Ioannidis, J. P. A., Bendavid, E., Salholz-Hillel, M., et al. 2022. Massive covidization of research citations and the citation elite. Proceedings of the National Academy of Sciences, 119, 28 e2204074119. https://doi.org/10.1073/pnas.2204074119

Ioannidis, J. P. A., Tarone, R. E., & McLaughlin, J. K. (2011). The false-positive to false-negative ratio in epidemiologic studies. Epidemiology, 22, 450–456. http://doi.org/10.1097/EDE.0b013e31821b506e

Iqbal, S. A., Wallach, J. D., Khoury, M. J., et al. (2016). Reproducible research practices and transparency across the biomedical literature. PLoS Biology, 4(1), e1002333. http://doi.org/10.1371/journal.pbio.1002333

Jefferson, T., Del Mar, C. B., Dooley, L., et al. (2020). Physical interventions to interrupt or reduce the spread of respiratory viruses. Cochrane Database of Systematic Reviews, 11(11), CD006207. http://doi.org/10.1002/14651858.CD006207.pub5

Jenson, H. B. (2020). How did "flatten the curve" become "flatten the economy?" A perspective from the United States of America. Asian Journal of Psychiatry, 51, 102165. http://doi.org/10.1016/j.ajp.2020.102165

Kavvoura, F. K., Liberopoulos, G., & Ioannidis, J. P. (2007). Selection in reported epidemiological risks: an empirical assessment. PLoS Medicine, 4(3), e79. http://doi.org/10.1371/journal.pmed.0040079

Keown, S. (2012). Biases Rife in Research, Ioannidis Says. NIH Record, LXIV(10). https://nihrecord.nih.gov/sites/recordNIH/files/pdf/2012/NIH-Record-2012-05-11.pdf

Kim, M. S., Seong, D., Li, H., et al. (2022). Comparative effectiveness of N95, surgical or medical, and non-medical facemasks in protection against respiratory virus infection: A systematic review and network meta-analysis. Reviews in Medical Virology, 32(5), e2336. https://doi.org/10.1002/rmv.2336

Kindzierski, W., Young, S., Meyer, T., et al. (2021). Evaluation of a meta-analysis of ambient air quality as a risk factor for asthma exacerbation. Journal of Respiration, 1(3), 173−196. https://doi.org/10.3390/jor1030017

Kinsella, C. M., Santos, P. D., Postigo-Hidalgo, I., et al. (2020). Preparedness needs research: How fundamental science and international collaboration accelerated the response to COVID-19. PLoS Pathogens, 16(10): e1008902. https://doi.org/10.1371/journal.ppat.1008902

Kosnik, L. R., & Bellas, A. (2020). Drivers of COVID-19 stay at home orders: Epidemiologic, economic, or political concerns? Economics of Disasters and Climate Change, 4(3), 503–514. https://doi.org/10.1007/s41885-020-00073-0

Kupferschmidt, K. (2022). WHO's departing chief scientist regrets errors in debate over whether SARS-CoV-2 spreads through air. Science, 22 November 2022. https://doi.org/10.1126/science.adf9731

Landis, S. C., Amara, S. G., Asadullah, K., et al. (2012). A call for transparent reporting to optimize the predictive value of preclinical research. Nature, 490(7419), 187–191. https://doi.org/10.1038/nature11556

Lavezzo, E., Franchin, E., Ciavarella, C., et al. (2020). Suppression of a SARS-CoV-2 outbreak in the Italian municipality of Vo. Nature, 584, 425–429. https://doi.org/10.1038/s41586-020-2488-1

Liu, I. T., Prasad, V, & Darrow, J. J. (2021). Evidence for Community Cloth Face Masking to Limit the Spread of SARS-CoV-2: A Critical Review. CATO Working Paper No. 64. November 8, 2021. The CATO Institute, Washington, DC. https://www.cato.org/sites/cato.org/files/2021-11/working-paper-64.pdf

Marcon, A., Nguyen, G., Rava, M., et al. (2015). A score for measuring health risk perception in environmental surveys. Science of the Total Environment, 527-528, 270–278. https://doi.org/10.1016/j.scitotenv.2015.04.110

Magness P. (2021). The Failures of Pandemic Central Planning. October 1, 2021. http://doi.org/10.2139/ssrn.3934452

Matrajt, L., & Leung, T. (2020). Evaluating the effectiveness of social distancing interventions to delay or flatten the epidemic curve of Coronavirus disease. Emerging Infectious Diseases, 26(8), 1740−1748. https://doi.org/10.3201/eid2608.201093

Members, W.-C. J. M. (2020). Report of the WHO-China Joint Mission on Coronavirus Disease 2019 (COVID-19). World Health Organization (WHO). https://www.who.int/docs/default-source/coronaviruse/who-china-joint-mission-on-covid-19-final-report.pdf

Menter, T., Haslbauer, J. D., Nienhold, R., et al. (2020). Postmortem examination of COVID-19 patients reveals diffuse alveolar damage with severe capillary congestion and variegated findings in lungs and other organs suggesting vascular dysfunction. Histopathology, 77(2), 198−209. https://doi.org/10.1111/his.14134

Meyerowitz., E. A., Richterman, A., Gandhi, R. T., et al. (2021). Transmission of SARS-CoV-2: A Review of Viral, Host, and Environmental Factors. Annals of Internal Medicine, 174(1), 69−79. https://doi.org/10.7326/M20-5008

Michaud, D. S., Feder, K., Voicescu, S. A. (2018). Clarifications on the design and interpretation of conclusions from Health Canada's study on wind turbine noise and health. Acoustics Australia, 46, 99–110. https://doi.org/10.1007/s40857-017-0125-4

Miller, I. (2022). Unmasked: The Global Failure of COVID Mask Mandates. Post Hill Press: Brentwood, TN.

Moffatt, S., & Bhopal, R. (2000). Study on environmental hazards is flawed. British Medical Journal, 320(7244), 1274. https://doi.org/10.1136/bmj.320.7244.1274

Moffatt S., Mulloli T.P., Bhopal R., et al. (2000). An exploration of awareness bias in two environmental epidemiology studies. Epidemiology, 11(2), 199−208. https://doi.org/10.1097/00001648-200003000-00020

Moher, D., Liberati, A., Tetzlaff, J., et al.; The PRISMA Group. (2009). Preferred reporting items for systematic reviews and meta-analyses: the PRISMA statement. PLoS Medicine, 6(7), e1000097. https://doi.org/10.1371/journal.pmed.1000097

Moonesinghe, R., Khoury, M. J., & Janssens, A. C. J. W. (2007). Most published research findings are false—But a little replication goes a long way. PLoS Medicine, 4, e28. https://doi.org/10.1146/10.1371/journal.pmed.0040028

Mosley, V., & Wyckoff, R. (1946). Electron micrography of the virus of influenza. Nature, 157, 263. https://doi.org/10.1038/157263a0

Munafo, M. R., Nosek, B. A., Bishop, D. V. M., et al. (2017). A manifesto for reproducible science. Nature Human Behaviour, 1, 0021. https://doi.org/10.1038/s41562-016-0021

Murad, M. H., Asi, N., Alsawas, M., et al. (2016). New evidence pyramid. BMJ Evidence-Based Medicine, 21(4), 125−127. http://dx.doi.org/10.1136/ebmed-2016-110401

Nanda, A., Hung, I., Kwong, A., et al. (2021). Efficacy of surgical masks or cloth masks in the prevention of viral transmission: Systematic review, meta-analysis, and proposal for future trial. Journal of Evidence-Based Medicine, 14(2), 97–111. https://doi.org/10.1111/jebm.12424

National Institutes of Health (NIH). 2009. Understanding a common cold virus. National Institutes of Health, Bethesda, MD. https://www.nih.gov/news-events/nih-research-matters/understanding-common-cold-virus

National Institutes of Health (NIH). 2017. Influenza virus biology. NIH Influenza Virus Resource help center, National Library of Medicine, Bethesda, MD. https://www.ncbi.nlm.nih.gov/genome/viruses/variation/help/flu-help-center/influenza-virus-biology/

Nissen, S. B., Magidson, T., Gross, K., et al. (2016). Publication bias and the canonization of false facts. eLife, 5, e21451. https://doi.org/10.7554/elife.21451

O'Conner, D. S., Green, S., & Higgins, J. P. T. (2008). Cochrane Handbook of Systematic Reviews of Intervention. (2008). Chichester, UK: Wiley-Blackwell.

Ollila, H. M., Laine, L., Koskela, J., et al. (2020). Systematic review and meta-analysis to examine the use of face mask intervention in mitigating the risk of spread of respiratory infections and if the effect of face mask use differs in different exposure settings and age groups. PROSPERO 2020 CRD42020205523. https://www.crd.york.ac.uk/prospero/display_record.php?ID=CRD42020205523

Ollila, H. M., Partinen, M., Koskela, J., et al. (2022). Face masks to prevent transmission of respiratory infections: Systematic review and meta-analysis of randomized controlled trials on face mask use. PLoS One. 2022 Dec 1;17(12):e0271517. doi: 10.1371/journal.pone.0271517

Paez, A. (2021). Reproducibility of research during COVID-19: Examining the case of population density and the basic reproductive rate from the perspective of spatial analysis. Geographical Analysis, 54, 860–880. https://doi.org/10.1111/gean.12307

Patel, A., & Jernigan, D. B. (2020). 2019-nCoV CDC Response Team. Initial public health response and interim clinical guidance for the 2019 novel coronavirus outbreak—United States, December 31, 2019–February 4, 2020. Morbidity and Mortality Weekly Report (MMWR), 69(5), 140–146. https://doi.org/10.15585/mmwr.mm6905e1

Peng, R. D., & Hicks, S. C. (2021). Reproducible research: A retrospective. Annual Review of Public Health, 42, 79−93. https://doi.org/10.1146/annurev-publhealth-012420-105110

Prather, K. A., Wang, C. C., & Schooley, R. T. (2020). Reducing transmission of SARS-CoV-2, Science, 368(6498), 1422–1424. https://doi.org/10.1126/science.abc6197

Rabinowitz, P. M., Slizovskiy, I. B., Lamers, V., et al. (2015). Proximity to natural gas wells and reported health status: results of a household survey in Washington County, Pennsylvania. Environmental Health Perspectives, 123(1), 21–26. https://doi.org/10.1289/ehp.1307732

Randall, D., & Welser, C. (2018). The Irreproducibility Crisis of Modern Science: Causes, Consequences, and the Road to Reform. New York, NY: National Association of Scholars. Nas.org/reports/the-irreproducibility-crisis-of-modern-science

Scheuch, G. (2020). Breathing is enough: For the spread of Influenza virus and SARS-CoV-2 by breathing only. Journal of Aerosol Medicine and Pulmonary Drug Delivery, 33(4), 230−234. https://doi.org/10.1089/jamp.2020.1616

Seong, D., Shin, J., & Kim, M. (2020). Comparative efficacy of N95, surgical, medical, and non-medical facemasks in respiratory virus transmission prevention: a living systematic review and network meta-analysis. PROSPERO 2020 CRD42020214729. https://www.crd.york.ac.uk/prospero/display_record.php?ID=CRD42020214729

Shusterman, D. (1992). Critical review: The health significance of environmental odor pollution. Archives of Environmental & Occupational Health, 47, 76–87. https://doi.org/10.1080/00039896.1992.9935948

Schweder, T., & Spjøtvoll, E. (1982). Plots of p-values to evaluate many tests simultaneously. Biometrika, 69, 493−502. https://doi.org/10.1093/biomet/69.3.493

Smith-Sivertsen, T., Tchachtchine, V., & Lund, E. (2000). Self-reported airway symptoms in a population exposed to heavy industrial pollution: What is the role of public awareness? Epidemiology, 11(6), 739−740. https://doi.org/10.1097/00001648-200011000-00027.

Stanley, W. M. (1944). The size of the Influenza virus. Journal of Experimental Medicine, 79(3), 267–283. https://doi.org/10.1084/jem.79.3.267

Stodden, V., Seiler, J., & Ma, Z. K. (2018). An empirical analysis of journal policy effectiveness for computational reproducibility. Proceedings of the National Academy of Sciences, 115, 2584–2589. https://doi.org/10.1073/pnas.1708290115

Stott, E. J., & Killington, R. A. (1972). Rhinoviruses. Annual Review of Microbiology, 26(1), 503−524. https://doi.org/10.1146/annurev.mi.26.100172.002443

Sumner, J., Haynes L., Nathan S., et al. (2020). Reproducibility and reporting practices in COVID-19 preprint manuscripts. medRxiv, 2020.03.24.20042796. https://doi.org/10.1101/2020.03.24.20042796

Tellier, R. (2006). Review of aerosol transmission of Influenza A virus. Emerging Infectious Diseases, 12(11), 1657–1662. https://doi.org/10.3201/eid1211.060426

Tellier, R. (2009). Aerosol transmission of Influenza A virus: a review of new studies. Journal of the Royal Society Interface, 6, S783–S790. https://doi.org/10.1098/rsif.2009.0302.focus

Tran, T. Q., Mostafa, E. M., Ravikulan, R., et al. (2020). Efficacy of facemasks against airborne infectious diseases: a systematic review and network meta-analysis of randomized-controlled trials. PROSPERO 2020 CRD42020178516. https://www.crd.york.ac.uk/prospero/display_record.php?ID=CRD42020178516

Tran, T. Q., Mostafa, E. M., Tawfik, G. M., et al. (2021). Efficacy of face masks against respiratory infectious diseases: A systematic review and network analysis of randomized-controlled trials. Journal of Breath Research, 15, 047102. https://doi.org/10.1088/1752-7163/ac1ea5

University of Maryland (UMD). (2022). Global COVID-19 Trends and Impact Survey. Joint Program in Survey Methodology, University of Maryland, College Park, MD. https://jpsm.umd.edu/landingtopic/global-covid-19-trends-and-impact-survey

Villeneuve, P. J., Ali, A., Challacombe, L., et al. (2009). Intensive hog farming operations and self-reported health among nearby rural residents in Ottawa, Canada. BMC Public Health, 9, 330. https://doi.org/10.1186/1471-2458-9-330

Wang, C. C., Prather, K. A., Sznitman, J., et al. (2021). Airborne transmission of respiratory viruses. Science, 373(6558), eabd9149. https://doi.org/10.1126/science.abd9149

Ware, J. J., & Munafo, M. R. (2015). Significance chasing in research practice: Causes, consequences and possible solutions. Addiction, 110(1), 4–8. https://doi.org/10.1111/add.12673

World Health Organization (WHO). (2019). Non-pharmaceutical public health measures for mitigating the risk and impact of epidemic and pandemic influenza; Annex: Report of systematic literature reviews. No. WHO/WHE/IHM/GIP/2019.1. WHO, Geneva. https://apps.who.int/iris/bitstream/handle/10665/329439/WHO-WHE-IHM-GIP-2019.1-eng.pdf

Xiao, J., Shiu, E. Y. C., Gao, H., et al. (2020). Nonpharmaceutical measures for pandemic influenza in nonhealthcare settings - Personal protective and environmental measures. Emerging Infectious Diseases, 26(5), 967–975. https://doi.org/10.3201/eid2605.190994

Young, S. S., & Kindzierski, W. B. (2019). Evaluation of a meta-analysis of air quality and heart attacks, a case study. Critical Reviews in Toxicology, 49(1), 85–94. https://doi.org/10.1080/10408444.2019.1576587

Young, S.S., Cheng, K.-C., Chen, J. H., et al. (2022). Reliability of a meta-analysis of air quality−asthma cohort studies. International Journal of Statistics and Probability, 11(2), 61−76. https://doi.org/10.5539/ijspv11n2p61

Young, S. S., & Kindzierski, W. B. (2022). Statistical reliability of a diet-disease association meta-analysis. International Journal of Statistics and Probability, 11(3), 40–50. https://doi.org/10.5539/ijsp.v11n3p40

Young, S. S., & Kindzierski, W. B. (2023). Reproducibility of health claims in meta-analysis studies of COVID quarantine (stay-at-home) orders. International Journal of Statistics and Probability, 12(1), 54–65. https://doi.org/10.5539/ijsp.v12n1p54

Zhu, N., Zhang, D., Wang, W., et al. (2020). China Novel Coronavirus Investigating and Research Team: A novel coronavirus from patients with pneumonia in China, 2019. New England Journal of Medicine, 382, 727–733. https://doi.org/10.1056/NEJMoa2001017

**APPENDICES**

Appendix A –    P-value plots and summary statistics of data sets for 'plausible true null' and 'plausible true alternative' hypothesis outcomes

Appendix B –    Cochrane Central Register of Controlled Trials (CENTRAL) and PubMed search results

Appendix C –    Results for other meta-analyses or systematic reviews

**Appendix A – P-value plots and summary statistics of data sets for
'plausible true null' and 'plausible true alternative' hypothesis outcomes**

Figure A1. p-value plots for meta-analysis of small observational datasets representing: (i) plausible true null hypothesis for a petroleum refinery worker−chronic myeloid leukemia causal relationship (n=12) after Schnatter et al. (2018) and (ii) plausible true alternative hypothesis for a petroleum refinery worker−mesothelioma causal relationship (n=10) after Schnatter et al. (2018).

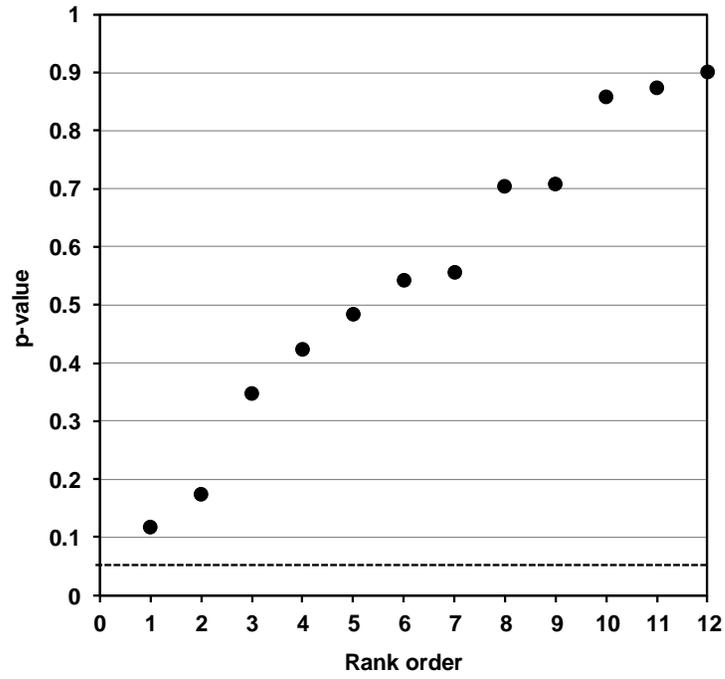

(i)

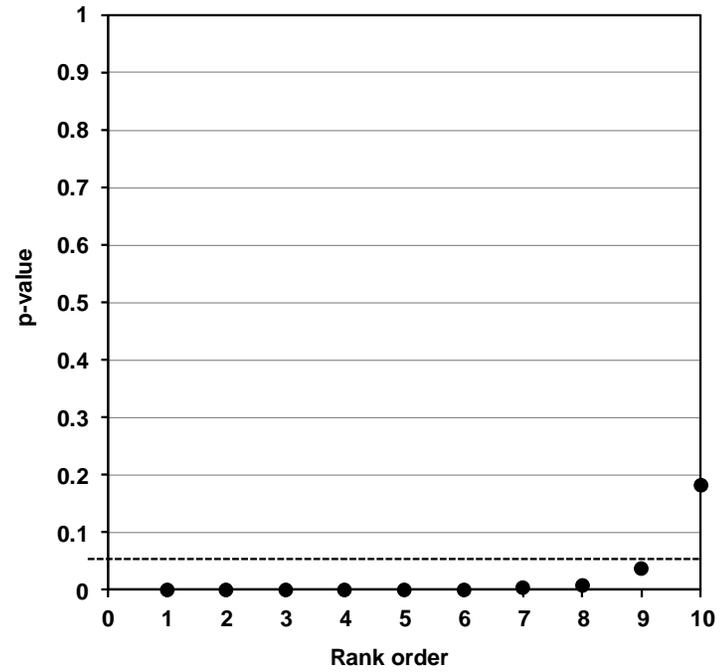

(ii)

Figure A2. p-value plots for meta-analysis of large observational datasets representing: (i) plausible true null hypothesis for an elderly long-term exercise training−mortality & morbidity causal relationship (n=69) after de Souto Barreto et al. (2019) and (ii) plausible true alternative hypothesis for a smoking−lung squamous cell carcinoma causal relationship (n=102) after Lee et al. (2012).

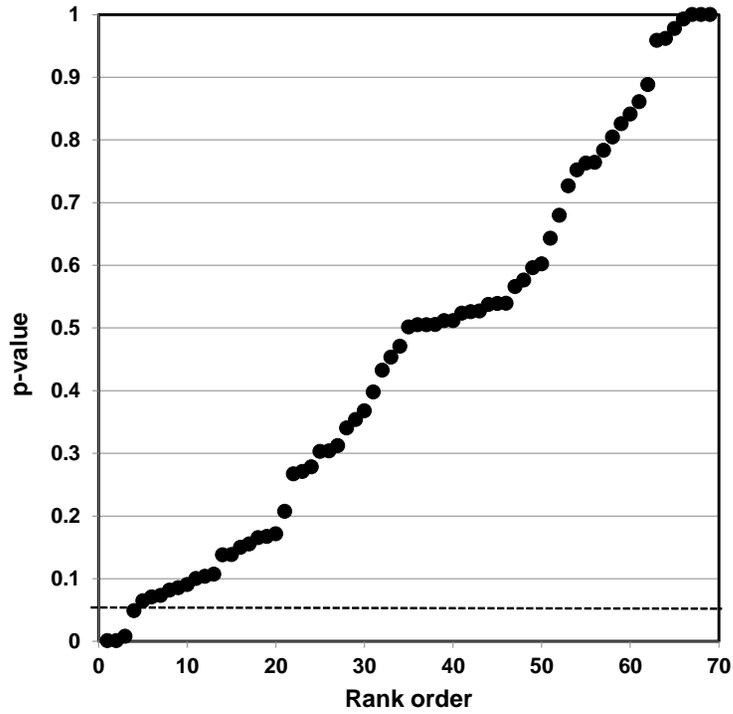

(i)

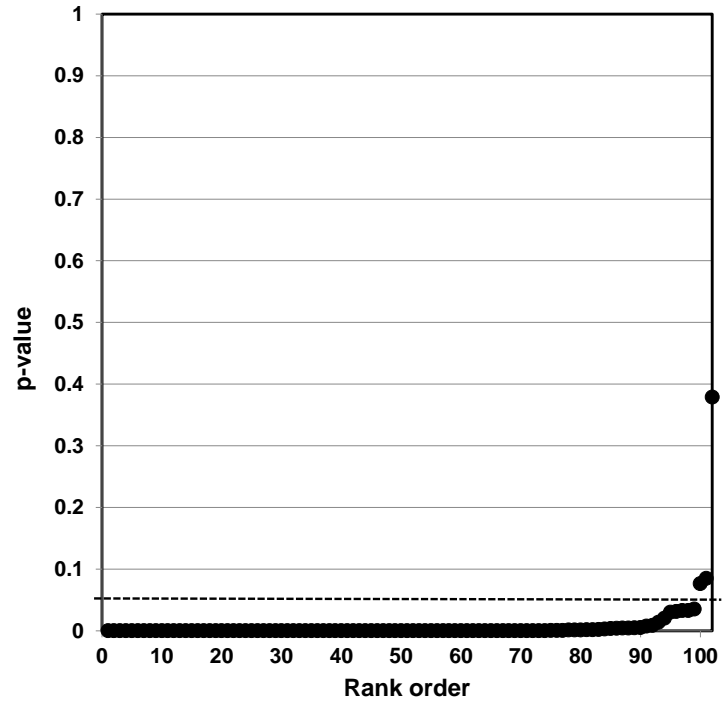

(ii)

**Explanation of Figures A1 and A1**

Figure A1 presents a set of p-value plots for small meta-analysis datasets (i.e., n<15 base papers) showing plausible true null and true alternative hypothesis representing selected cancers in petroleum refinery workers after Schnatter et al. (2018):

- Figure A1 (i); left image – presents p-values as a sloped line from left to right at approximately 45-degrees representing a plausible true null hypothesis for a chronic myeloid leukemia causal relationship in petroleum refinery workers (n=12) after Schnatter et al. (2018).
- Figure A1 (ii); right image – presents a majority of p-values below the .05 line representing a plausible true alternative hypothesis for a mesothelioma causal relationship in petroleum refinery workers (n=10) after Schnatter et al. (2018).

Figure A2 presents a set of p-value plots for large meta-analysis datasets (i.e., n>65 base papers) showing plausible true null and true alternative hypothesis:

- Figure A2 (i); left image – presents p-values as a sloped line from left to right at approximately 45-degrees representing a plausible true null hypothesis for an elderly long-term exercise training−mortality & morbidity causal relationship (n=69) after de Souto Barreto et al. (2019).
- Figure A2 (ii); left image – presents a majority of p-values below the .05 line representing a plausible true alternative hypothesis for a smoking−lung squamous cell carcinoma causal relationship (n=102) after Lee et al. (2012).

**Summary statistics of data sets used in the p-value plots**

**Meta-analysis of selected cancers in petroleum refinery workers** (after Schnatter et al., 2018^)

Note: Base study=base study 1st author name in Schnatter et al. (2018); RR=relative risk; LCL=lower confidence limit; UCL=upper confidence limit; bold, italicized p-value <0.05; p-values were calculated using the method of Altman, D. G., & Bland, J. M. (2011). How to obtain the p-value from a confidence interval. British Medical Journal, 343, d2304. https://doi.org/10.1136/bmj.d2304. A p-value calculated as ≤0.0001 was recorded as 0.0001.

Chronic myeloid leukemia risk for petroleum refinery workers:

| Base study | RR | LCL | UCL | p-value |
|------------|------|------|------|---------|
| Collingwood 1996 | 0.53 | 0.07 | 3.74 | 0.54263 |
| Divine 1999a | 1.05 | 0.60 | 1.85 | 0.87478 |
| Gun 2006b | 1.09 | 0.45 | 2.61 | 0.85800 |
| Huebner 2004 | 1.68 | 0.88 | 3.23 | 0.11778 |
| Lewis 2000a | 1.08 | 0.35 | 3.35 | 0.90196 |
| Rushton 1993a | 0.89 | 0.50 | 1.61 | 0.70923 |
| Satin 1996 | 0.85 | 0.38 | 1.88 | 0.70346 |
| Satin 2002 | 0.45 | 0.14 | 1.39 | 0.17355 |
| Tsai 2007 | 0.66 | 0.21 | 2.05 | 0.48425 |
| Wong 2001a | 1.31 | 0.55 | 3.15 | 0.55549 |
| Wong 2001b | 1.96 | 0.49 | 7.84 | 0.34689 |
| Wongsrichanalia 1989 | 0.44 | 0.06 | 3.12 | 0.42318 |

Median RR = 0.97 (~1); range of the RR IQR (Interquartile Range) = 0.63−1.15

Mesothelioma risk for petroleum refinery workers (based on mesothelioma subgroup analysis using Schnatter et al. (2018) 'Best Methods' dataset):

| Base study | RR | LCL | UCL | p-value |
|------------|------|------|-------|---------|
| Devine 1999a | 2.97 | 2.21 | 3.99 | ***0.0001*** |
| Gamble 2000 | 2.43 | 1.35 | 4.39 | ***0.00321*** |
| Gun 2006a | 3.77 | 2.14 | 6.64 | ***0.0001*** |
| Honda 1995 | 2.00 | 1.04 | 3.84 | ***0.03720*** |
| Hornstra 1993 | 5.51 | 3.38 | 8.99 | ***0.0001*** |
| Huebner 2009 | 2.44 | 1.83 | 3.24 | ***0.0001*** |
| Kaplan 1986 | 2.41 | 1.26 | 4.64 | ***0.00817*** |
| Lewis 2000a | 8.68 | 5.77 | 13.06 | ***0.0001*** |
| Tsai 2003 | 2.16 | 0.70 | 6.69 | 0.18215 |
| Tsai 2007 | 2.50 | 1.63 | 3.83 | ***0.0001*** |

Median RR = 2.47 (>2); range of the RR IQR = 2.42−3.57

^  Schnatter, A. R., Chen, M., DeVilbiss, E. A., Lewis, R. J., & Gallagher, E. M. (2018). Systematic review and meta-analysis of selected cancers in petroleum refinery workers. Journal of Occupational and Environmental Medicine, 60(7), e329−e342. https://doi.org/10.1097/JOM.0000000000001336

**Meta-analysis of elderly long-term exercise training−mortality & morbidity risk** (after de Souto Barreto et al., 2019*)

Note: Base study=base study 1st author name in de Souto Barreto et al. (2019); RR=relative risk; LCL=lower confidence limit; UCL=upper confidence limit; bold, italicized p-value <0.05; p-values were calculated using the method of Altman, D. G., & Bland, J. M. (2011). How to obtain the p-value from a confidence interval. British Medical Journal, 343, d2304. https://doi.org/10.1136/bmj.d2304. A p-value calculated as ≤0.0001 was recorded as 0.0001.

| Outcome | No. | Base Study ID (n=69) | RR | LCL | UCL | *p-value* |
|---|---|---|---|---|---|---|
| Mortality | 1 | Belardinelli et al. 2012 | 0.38 | 0.13 | 1.15 | 0.08153 |
| | 2 | Barnett et al. 2003 | 0.14 | 0.01 | 2.63 | 0.16736 |
| | 3 | O'Connor et al. 2009 | 0.96 | 0.80 | 1.16 | 0.67980 |
| | 4 | Campbell et al. 1997 | 0.50 | 0.09 | 2.70 | 0.43245 |
| | 5 | El−Khoury et al. 2015 | 0.84 | 0.26 | 2.72 | 0.78350 |
| | 6 | Galvão et al. 2014 | 3.00 | 0.13 | 71.92 | 0.50544 |
| | 7 | Gianoudis et al. 2014 | 1.00 | 0.06 | 15.72 | 1.00000 |
| | 8 | Hewitt et al. 2018 | 1.02 | 0.52 | 2.03 | 0.95866 |
| | 9 | Karinkanta et al. 2007 | 0.33 | 0.01 | 7.93 | 0.52568 |
| | 10 | Kemmler et al. 2010 | 0.33 | 0.01 | 8.10 | 0.52704 |
| | 11 | King et al. 2002 | 0.32 | 0.01 | 7.68 | 0.51168 |
| | 12 | Kovács et al. 2013 | 0.40 | 0.14 | 1.17 | 0.09039 |
| | 13 | Lam et al. 2012 | 0.64 | 0.06 | 7.05 | 0.72673 |
| | 14 | Lam et al. 2015 | 0.30 | 0.03 | 2.82 | 0.30309 |
| | 15 | Lord et al. 2003 | 4.84 | 0.55 | 42.33 | 0.15519 |
| | 16 | Merom et al. 2015 | 1.36 | 0.22 | 8.23 | 0.75225 |
| | 17 | Pahor et al. 2006 | 0.99 | 0.14 | 6.97 | 0.99276 |
| | 18 | Pahor et al. 2014/Gill et al. 2016 | 1.14 | 0.76 | 1.71 | 0.53739 |
| | 19 | Patil et al. 2015 | 0.11 | 0.01 | 2.04 | 0.10355 |
| | 20 | Pitkälä et al. 2013 | 0.25 | 0.06 | 1.14 | 0.06455 |
| | 21 | Prescott et al. 2008 | 0.42 | 0.08 | 2.12 | 0.30363 |
| | 22 | Rejeski et al. 2017 | 0.34 | 0.01 | 8.16 | 0.53914 |
| | 23 | Rolland et al. 2007 | 0.88 | 0.34 | 2.28 | 0.80441 |
| | 24 | Sherrington et al. 2014 | 1.10 | 0.46 | 2.63 | 0.84135 |
| | 25 | Underwood et al. 2013 | 1.06 | 0.84 | 1.35 | 0.64301 |
| | 26 | Van Uffelen et al. 2008 | 0.36 | 0.01 | 8.72 | 0.56573 |
| | 27 | von Stengel et al. 2011 | 0.34 | 0.01 | 8.15 | 0.53907 |
| | 28 | Voukelatos et al. 2015 | 9.09 | 0.49 | 167.75 | 0.13843 |
| | 29 | Wolf et al. 2003 | 0.97 | 0.14 | 6.86 | 0.97786 |
| Hospitalization | 30 | Belardinelli et al. 2012 | 0.30 | 0.15 | 0.62 | ***0.00092*** |
| | 31 | O'Connor et al. 2009 | 0.97 | 0.91 | 1.03 | 0.34039 |
| | 32 | Hambrecht et al. 2004 | 0.16 | 0.02 | 1.31 | 0.08551 |
| | 33 | Hewitt et al. 2018 | 0.64 | 0.27 | 1.50 | 0.31209 |
| | 34 | Kovács et al. 2013 | 2.00 | 0.19 | 21.21 | 0.57619 |
| | 35 | Messier et al. 2013 | 8.54 | 0.46 | 157.06 | 0.14992 |
| | 36 | Mustata et al. 2011 | 0.33 | 0.02 | 7.32 | 0.47075 |

| Outcome | No. | Base Study ID (n=69) | RR | LCL | UCL | *p-value* |
|---------|-----|----------------------|----|-----|-----|-----------|
| Hospitalization | 37 | Pahor et al. 2006 | 0.99 | 0.68 | 1.44 | 0.96195 |
| | 38 | Pahor et al. 2014/Gill et al. 2016 | 1.10 | 0.99 | 1.22 | 0.07332 |
| | 39 | Pitkala et al. 2013 | 0.78 | 0.55 | 1.12 | 0.17166 |
| | 40 | Rejeski et al. 2017 | 3.04 | 0.13 | 73.46 | 0.50161 |
| | 41 | Rolland et al. 2007 | 1.82 | 0.95 | 3.49 | 0.07083 |
| Injurious falls | 42 | Barnett et al. 2003 | 0.77 | 0.48 | 1.21 | 0.27108 |
| | 43 | Campbell et al. 1997 | 0.67 | 0.45 | 1.00 | ***0.04892*** |
| | 44 | El−Khoury et al. 2015 | 0.90 | 0.78 | 1.05 | 0.16541 |
| | 45 | Hewitt et al. 2018 | 0.58 | 0.42 | 0.81 | ***0.00120*** |
| | 46 | MacRae et al. 1994 | 0.16 | 0.01 | 2.92 | 0.20731 |
| | 47 | Pahor et al. 2014/Gill et al. 2016 | 0.89 | 0.66 | 1.20 | 0.45350 |
| | 48 | Patil et al. 2015 | 0.51 | 0.31 | 0.84 | ***0.00810*** |
| | 49 | Pitkälä et al. 2013 | 0.65 | 0.39 | 1.09 | 0.10016 |
| | 50 | Reinsch et al. 1992 | 1.46 | 0.37 | 5.81 | 0.60232 |
| Fractures | 51 | Belardinelli et al. 2012 | 0.19 | 0.01 | 3.89 | 0.27847 |
| | 52 | O'Connor et al. 2009 | 0.60 | 0.32 | 1.11 | 0.10725 |
| | 53 | El−Khoury et al. 2015 | 0.88 | 0.60 | 1.25 | 0.50488 |
| | 54 | Gianoudis et al. 2014 | 3.00 | 0.12 | 72.57 | 0.51161 |
| | 55 | Hewitt et al. 2018 | 0.80 | 0.20 | 3.11 | 0.76275 |
| | 56 | Karinkanta et al. 2007 | 1.00 | 0.15 | 6.73 | 1.00000 |
| | 57 | Kemmler et al. 2010 | 0.49 | 0.19 | 1.25 | 0.13795 |
| | 58 | Kovács et al. 2013 | 3.00 | 0.13 | 71.56 | 0.50509 |
| | 59 | Lam et al. 2012 | 1.27 | 0.06 | 28.95 | 0.88844 |
| | 60 | Pahor et al. 2014/Gil et al. 2016 | 0.87 | 0.63 | 1.19 | 0.39774 |
| | 61 | Patil et al. 2015 | 0.66 | 0.28 | 1.59 | 0.35403 |
| | 62 | Pitkälä et al. 2013 | 1.00 | 0.26 | 3.84 | 1.00000 |
| | 63 | Reinsch et al. 1992 | 0.45 | 0.04 | 4.78 | 0.52344 |
| | 64 | Rolland et al. 2007 | 2.50 | 0.50 | 12.44 | 0.26692 |
| | 65 | Sherrington et al. 2014 | 0.92 | 0.46 | 1.85 | 0.82585 |
| | 66 | Underwood et al. 2013 | 1.05 | 0.63 | 1.74 | 0.86094 |
| | 67 | Villareal et al. 2011 | 0.52 | 0.05 | 5.39 | 0.59601 |
| | 68 | von Stengel et al. 2011 | 0.58 | 0.18 | 1.87 | 0.36778 |
| | 69 | Wolf et al. 2003 | 0.78 | 0.17 | 3.67 | 0.76405 |

Median RR = 0.80 (<1); range of the RR IQR = 0.42−1.05

**Meta-analysis of smoking−lung squamous cell carcinoma risk** (after Lee et al., 2012⁰)

Note: Base study=base study 1ˢᵗ author name in Lee et al. (2012); RR=relative risk; LCL=lower confidence limit; UCL=upper confidence limit; p-value calculated after Altman (2011); bold, italicized p-value <0.05; p-values were calculated using the method of Altman, D. G., & Bland, J. M. (2011). How to obtain the p-value from a confidence interval. British Medical Journal, 343, d2304. https://doi.org/10.1136/bmj.d2304. A p-value calculated as ≤0.0001 was recorded as 0.0001.

| Place | No. | Base Study ID (n=102) | RR | LCL | UCL | *p-value* |
|-------|-----|----------------------|-----|-----|-----|-----------|
| USA | 1 | 1948 WYNDE4 m | 12.79 | 6.19 | 26.14 | ***0.0001*** |
| | 2 | 1948 WYNDE4 f | 2.82 | 2.55 | 13.31 | ***0.01380*** |
| | 3 | 1949 BRESLO c | 3.69 | 2.06 | 6.62 | ***0.0001*** |
| | 4 | 1952 HAMMON m | 16.88 | 6.29 | 45.29 | ***0.0001*** |
| | 5 | 1955 HAENSZ f | 3.00 | 1.90 | 4.73 | ***0.0001*** |
| | 6 | 1957 BYERS1 m | 8.29 | 5.29 | 13.00 | ***0.0001*** |
| | 7 | 1960 LOMBA2 f | 4.24 | 2.40 | 7.50 | ***0.0001*** |
| | 8 | 1962 WYNDE2 m | 19.72 | 6.21 | 62.59 | ***0.0001*** |
| | 9 | 1964 OSANN2 f | 35.10 | 4.80 | 256.00 | ***0.00048*** |
| | 10 | 1966 WYNDE3 m | 18.29 | 5.71 | 58.56 | ***0.0001*** |
| | 11 | 1966 WYNDE3 f | 6.79 | 2.45 | 18.82 | ***0.00025*** |
| | 12 | 1968 HINDS f | 16.13 | 7.66 | 33.97 | ***0.0001*** |
| | 13 | 1969 STAYNE m | 3.47 | 2.17 | 5.56 | ***0.0001*** |
| | 14 | 1969 WYNDE6 m | 18.59 | 12.74 | 27.13 | ***0.0001*** |
| | 15 | 1969 WYNDE6 f | 32.37 | 17.66 | 59.35 | ***0.0001*** |
| | 16 | 1975 COMSTO m | 8.07 | 1.91 | 34.02 | ***0.00452*** |
| | 17 | 1975 COMSTO f | 46.20 | 2.74 | 778.83 | ***0.00784*** |
| | 18 | 1976 BUFFLE m | 14.03 | 4.73 | 41.61 | ***0.0001*** |
| | 19 | 1976 BUFFLE f | 13.04 | 3.99 | 42.66 | ***0.0001*** |
| | 20 | 1979 CORREA c | 28.30 | 18.60 | 43.20 | ***0.0001*** |
| | 21 | 1979 SIEMIA m | 22.70 | 6.90 | 75.20 | ***0.0001*** |
| | 22 | 1980 DORGAN m | 18.90 | 7.00 | 51.30 | ***0.0001*** |
| | 23 | 1980 DORGAN f | 11.10 | 7.20 | 17.10 | ***0.0001*** |
| | 24 | 1981 JAIN m | 18.00 | 5.50 | 111.00 | ***0.00018*** |
| | 25 | 1981 JAIN f | 25.50 | 7.93 | 156.00 | ***0.0001*** |
| | 26 | 1981 WU f | 24.29 | 3.40 | 173.76 | ***0.00153*** |
| | 27 | 1983 BAND m | 37.45 | 17.60 | 79.58 | ***0.0001*** |
| | 28 | 1984 BROWN2 m | 11.10 | 9.50 | 12.90 | ***0.0001*** |
| | 29 | 1984 BROWN2 f | 20.10 | 16.40 | 24.80 | ***0.0001*** |
| | 30 | 1984 OSANN m | 36.10 | 17.80 | 73.30 | ***0.0001*** |
| | 31 | 1984 OSANN f | 26.40 | 14.50 | 48.10 | ***0.0001*** |
| | 32 | 1984 SCHWAR m1 | 32.81 | 4.48 | 240.23 | ***0.0001*** |
| | 33 | 1984 SCHWAR m2 | 1.81 | 0.50 | 6.78 | 0.37881 |
| | 34 | 1984 SCHWAR f1 | 43.23 | 2.60 | 718.15 | ***0.00862*** |
| | 35 | 1984 SCHWAR f2 | 62.61 | 3.64 | 1076.10 | ***0.00441*** |
| | 36 | 1985 KHUDER m | 7.82 | 3.87 | 15.77 | ***0.0001*** |
| | 37 | 1986 ANDERS f | 25.57 | 10.29 | 63.56 | ***0.0001*** |
| | 38 | 1989 HEGMAN c | 30.80 | 12.48 | 76.03 | ***0.0001*** |

**(continued)**

| Place | No. | Base Study ID (n=102) | RR | LCL | UCL | *p-value* |
|-------|-----|----------------------|-----|-----|-----|-----------|
| Europe | 39 | 1947 ORMOS m | 10.14 | 2.41 | 42.79 | *0.00165* |
| | 40 | 1948 DOLL m | 13.17 | 4.12 | 42.10 | *0.0001* |
| | 41 | 1948 DOLL f | 2.13 | 1.06 | 4.27 | *0.03311* |
| | 42 | 1948 KREYBE m | 10.87 | 3.47 | 34.04 | *0.0001* |
| | 43 | 1948 KREYBE f | 2.29 | 0.89 | 5.88 | 0.08506 |
| | 44 | 1954 STASZE m | 57.77 | 3.58 | 933.17 | *0.00430* |
| | 45 | 1954 STASZE f | 32.45 | 1.32 | 800.04 | *0.03297* |
| | 46 | 1959 TIZZAN c | 2.70 | 1.99 | 3.67 | *0.0001* |
| | 47 | 1964 ENGELA m | 6.45 | 1.97 | 21.11 | *0.00211* |
| | 48 | 1966 TOKARS c | 6.80 | 1.20 | 38.70 | *0.03026* |
| | 49 | 1971 NOU m | 27.17 | 6.60 | 11.85 | *0.0001* |
| | 50 | 1971 NOU f | 7.09 | 1.35 | 37.19 | *0.02043* |
| | 51 | 1972 DAMBER m | 11.80 | 6.40 | 23.00 | *0.0001* |
| | 52 | 1975 ABRAHA m | 92.66 | 5.77 | 1488.21 | *0.00143* |
| | 53 | 1975 ABRAHA f | 5.35 | 2.22 | 12.90 | *0.00021* |
| | 54 | 1976 LUBIN2 m | 16.66 | 12.69 | 21.86 | *0.0001* |
| | 55 | 1976 LUBIN2 f | 5.78 | 4.34 | 7.71 | *0.0001* |
| | 56 | 1977 ALDERS m | 14.70 | 3.40 | 63.64 | *0.00035* |
| | 57 | 1977 ALDERS f | 6.09 | 2.68 | 13.82 | *0.0001* |
| | 58 | 1979 BARBON m | 14.52 | 6.35 | 33.20 | *0.0001* |
| | 59 | 1979 DOSEME m | 3.60 | 2.60 | 5.00 | *0.0001* |
| | 60 | 1980 JEDRYC m | 12.84 | 5.58 | 29.55 | *0.0001* |
| | 61 | 1983 SVENSS f | 12.62 | 3.97 | 40.14 | *0.0001* |
| | 62 | 1985 BECHER f | 10.69 | 2.43 | 47.00 | *0.00177* |
| | 63 | 1987 KATSOU f | 6.11 | 2.69 | 13.87 | *0.0001* |
| | 64 | 1988 JAHN m | 23.03 | 7.29 | 72.81 | *0.0001* |
| Asia | 65 | 1961 ISHIMA c | 21.00 | 3.38 | 868.40 | *0.03122* |
| | 66 | 1964 JUSSAW m | 25.43 | 13.87 | 46.63 | *0.0001* |
| | 67 | 1965 MATSUD m | 39.01 | 5.44 | 279.84 | *0.00029* |
| | 68 | 1976 CHAN m | 15.22 | 3.61 | 64.12 | *0.00023* |
| | 69 | 1976 CHAN f | 6.44 | 3.44 | 12.06 | *0.0001* |
| | 70 | 1976 LAMWK2 m | 6.89 | 2.65 | 17.90 | *0.0001* |
| | 71 | 1976 LAMWK2 f | 6.49 | 3.27 | 12.88 | *0.0001* |
| | 72 | 1976 TSUGAN m | 14.55 | 0.75 | 283.37 | 0.07657 |
| | 73 | 1978 ZHOU m | 3.14 | 1.90 | 5.18 | *0.0001* |
| | 74 | 1978 ZHOU f | 3.81 | 1.50 | 9.68 | *0.00496* |
| | 75 | 1981 KOO f | 4.15 | 2.46 | 6.98 | *0.0001* |
| | 76 | 1981 LAMWK f | 10.54 | 4.19 | 26.52 | *0.0001* |
| | 77 | 1981 XU3 m | 5.90 | 1.69 | 20.57 | *0.00540* |
| | 78 | 1981 XU3 f | 25.67 | 4.99 | 131.94 | *0.00012* |
| | 79 | 1982 ZHENG m | 16.82 | 6.05 | 46.71 | *0.0001* |
| | 80 | 1982 ZHENG f | 5.45 | 3.11 | 9.54 | *0.0001* |
| | 81 | 1983 LAMTH f | 8.10 | 4.16 | 15.77 | *0.0001* |

| Place | No. | Base Study ID (n=102) | RR | LCL | UCL | *p-value* |
|-------|-----|----------------------|-----|-----|-----|-----------|
| Asia | 82 | 1984 GAO m | 8.40 | 4.70 | 15.00 | *0.0001* |
| | 83 | 1984 GAO f | 7.20 | 4.60 | 11.10 | *0.0001* |
| | 84 | 1984 LUBIN m | 6.33 | 2.29 | 17.45 | *0.00040* |
| | 85 | 1985 CHOI m | 5.45 | 2.34 | 12.67 | *0.0001* |
| | 86 | 1985 CHOI f | 6.94 | 2.68 | 17.96 | *0.0001* |
| | 87 | 1985 WUWILL f | 4.20 | 3.00 | 5.90 | *0.0001* |
| | 88 | 1986 SOBUE m | 17.88 | 7.82 | 40.87 | *0.0001* |
| | 89 | 1986 SOBUE f | 8.74 | 5.09 | 15.02 | *0.0001* |
| | 90 | 1988 WAKAI m | 8.61 | 2.08 | 35.72 | *0.00305* |
| | 91 | 1988 WAKAI f | 25.23 | 6.87 | 92.66 | *0.0001* |
| | 92 | 1990 FAN c | 11.68 | 5.04 | 27.04 | *0.0001* |
| | 93 | 1990 GER c | 3.19 | 1.08 | 9.42 | *0.03547* |
| | 94 | 1990 LUO c | 10.90 | 2.50 | 47.90 | *0.00157* |
| | 95 | 1991 KIHARA c | 26.97 | 10.84 | 67.08 | *0.0001* |
| | 96 | 1997 SEOW f | 17.50 | 6.95 | 44.09 | *0.0001* |
| Other | 97 | 1978 JOLY m | 31.21 | 7.69 | 126.68 | *0.0001* |
| | 98 | 1978 JOLY f | 18.56 | 7.74 | 44.51 | *0.0001* |
| | 99 | 1987 PEZZOT m | 62.74 | 3.86 | 1019.50 | *0.00367* |
| | 100 | 1991 SUZUK2 c | 31.00 | 4.20 | 227.00 | *0.00078* |
| | 101 | 1993 DESTE2 m | 13.20 | 4.70 | 37.10 | *0.0001* |
| | 102 | 1994 MATOS m | 8.08 | 2.58 | 25.50 | *0.00038* |

Median RR = 12.8 (>2); range of the RR IQR = 6.6−24

[o]  Lee, P. N., Forey, B. A., & Coombs, K. J. (2012). Systematic review with meta-analysis of the epidemiological evidence in the 1900s relating smoking to lung cancer. BMC Cancer, 12, 385. https://doi.org/10.1186/1471-2407-12-385

**Appendix B – Cochrane Central Register of Controlled Trials (CENTRAL) and PubMed search results**

**Cochrane Central Register of Controlled Trials (CENTRAL) search results** (performed December 12, 2022)
Eligible studies that met search criteria: #11

1
**Universal screening for SARS-CoV-2 infection: a rapid review**
Meera Viswanathan, Leila Kahwati, Beate Jahn, et al.

2
**Antibody tests for identification of current and past infection with SARS-CoV-2**
Tilly Fox, Julia Geppert, Jacqueline Dinnes, et al., Cochrane COVID-19 Diagnostic Test Accuracy Group

3
**Rapid, point-of-care antigen tests for diagnosis of SARS-CoV-2 infection**
Jacqueline Dinnes, Pawana Sharma, Sarah Berhane, et al., Cochrane COVID-19 Diagnostic Test Accuracy Group

4
**SARS-CoV-2-neutralising monoclonal antibodies for treatment of COVID-19**
Nina Kreuzberger, Caroline Hirsch, Khai Li Chai, et al.

5
**Non-pharmacological measures implemented in the setting of long-term care facilities to prevent SARS-CoV-2 infections and their consequences: a rapid review**
Jan M Stratil, Renke L Biallas, Jacob Burns, et al.

6
**SARS-CoV-2-neutralising monoclonal antibodies to prevent COVID-19**
Caroline Hirsch, Yun Soo Park, Vanessa Piechotta, et al.

7
**Chloroquine or hydroxychloroquine for prevention and treatment of COVID-19**
Bhagteshwar Singh, Hannah Ryan, Tamara Kredo, et al.

8
**Remdesivir for the treatment of COVID-19**
Kelly Ansems, Felicitas Grundeis, Karolina Dahms, et al.

9
**Ivermectin for preventing and treating COVID-19**
Maria Popp, Stefanie Reis, Selina Schieber, et al.

10
**Measures implemented in the school setting to contain the COVID-19 pandemic**
Shari Krishnaratne, Hannah Littlecott, Kerstin Sell, et al.

11
**Physical interventions to interrupt or reduce the spread of respiratory viruses**
Tom Jefferson, Chris B Del Mar, Liz Dooley, et al.

12
**Colchicine for the treatment of COVID-19**
Agata Mikolajewska, Anna-Lena Fischer, Vanessa Piechotta, et al.

13
**Routine laboratory testing to determine if a patient has COVID-19**
Inge Stegeman, Eleanor A Ochodo, Fatuma Guleid, et al., Cochrane COVID-19 Diagnostic Test Accuracy Group

14
**Thoracic imaging tests for the diagnosis of COVID-19**
Sanam Ebrahimzadeh, Nayaar Islam, Haben Dawit, et al., Cochrane COVID-19 Diagnostic Test Accuracy Group

15
**Use of antimicrobial mouthwashes (gargling) and nasal sprays by healthcare workers to protect them when treating patients with suspected or confirmed COVID-19 infection**
Martin J Burton, Janet E Clarkson, Beatriz Goulao, et al.

16
**Janus kinase inhibitors for the treatment of COVID-19**
Andre Kramer, Carolin Prinz, Falk Fichtner, et al.

17
**Signs and symptoms to determine if a patient presenting in primary care or hospital outpatient settings has COVID-19**
Thomas Struyf, Jonathan J Deeks, Jacqueline Dinnes, et al., Cochrane COVID-19 Diagnostic Test Accuracy Group

18
**Convalescent plasma or hyperimmune immunoglobulin for people with COVID-19: a living systematic review**
Vanessa Piechotta, Claire Iannizzi, Khai Li Chai, et al.

19
**Anticoagulants for people hospitalised with COVID-19**
Ronald LG Flumignan, Vinicius T Civile, Jéssica Dantas de Sá Tinôco, et al.

20
**Digital contact tracing technologies in epidemics: a rapid review**
Andrew Anglemyer, Theresa HM Moore, Lisa Parker, et al.

21
**International travel-related control measures to contain the COVID-19 pandemic: a rapid review**
Jacob Burns, Ani Movsisyan, Jan M Stratil, et al.

22
**Quarantine alone or in combination with other public health measures to control COVID-19: a rapid review**
Barbara Nussbaumer-Streit, Verena Mayr, Andreea Iulia Dobrescu, et al.

23
**Interventions to support the resilience and mental health of frontline health and social care professionals during and after a disease outbreak, epidemic or pandemic: a mixed methods systematic review**
Alex Pollock, Pauline Campbell, Joshua Cheyne, et al.

24
**Personal protective equipment for preventing highly infectious diseases due to exposure to contaminated body fluids in healthcare staff**
Jos H Verbeek, Blair Rajamaki, Sharea Ijaz, et al.

25
**Antibiotics for the treatment of COVID-19**
Maria Popp, Miriam Stegemann, Manuel Riemer, et al.

26
**Barriers and facilitators to healthcare workers' adherence with infection prevention and control (IPC) guidelines for respiratory infectious diseases: a rapid qualitative evidence synthesis**

Catherine Houghton, Pauline Meskell, Hannah Delaney, et al.

Xiaomei Chen, Jiaojiao Jiang, Renjie Wang, et al.

Jerry S Zifodya, Jonah S Kreniske, Ian Schiller, et al.

53
**Xpert MTB/RIF Ultra and Xpert MTB/RIF assays for extrapulmonary tuberculosis and rifampicin resistance in adults**
Mikashmi Kohli, Ian Schiller, Nandini Dendukuri, et al.

54
**Rapid diagnostic tests for plague**
Sophie Jullien, Harsha A Dissanayake, Marty Chaplin

55
**Xpert MTB/RIF Ultra assay for tuberculosis disease and rifampicin resistance in children**
Alexander W Kay, Tara Ness, Sabine E Verkuijl, et al.

56
**Interleukin-1 blocking agents for treating COVID-19**
Mauricia Davidson, Sonia Menon, Anna Chaimani, et al.

57
**COVID-19 and its cardiovascular effects: a systematic review of prevalence studies**
Pierpaolo Pellicori, Gemina Doolub, Chih Mun Wong, et al.

58
**Interventions for the treatment of persistent post-COVID-19 olfactory dysfunction**
Lisa O'Byrne, Katie E Webster, Samuel MacKeith, et al.

59
**Interventions for the prevention of persistent post-COVID-19 olfactory dysfunction**
Katie E Webster, Lisa O'Byrne, Samuel MacKeith, et al.

60
**Systemic corticosteroids for the treatment of COVID-19: Equity-related analyses and update on evidence**
Carina Wagner, Mirko Griesel, Agata Mikolajewska, et al.

61
**Healthcare workers' perceptions and experiences of communicating with people over 50 years of age about vaccination: a qualitative evidence synthesis**
Claire Glenton, Benedicte Carlsen, Simon Lewin, et al.

**PubMed search results** (performed December 12, 2022)
Eligible studies that met search criteria: #'s 7, 13, 16, 18, 26, 52

1
Bundgaard H, Bundgaard JS, Raaschou-Pedersen DET, et al. Effectiveness of Adding a Mask Recommendation to Other Public Health Measures to Prevent SARS- CoV-2 Infection in Danish Mask Wearers: A Randomized Controlled Trial. Ann Intern Med. 2021 Mar;174(3):335-343. https://doi.org/10.7326/M20-6817

2
Chu DK, Akl EA, Duda S, Solo K, et al., COVID-19 Systematic Urgent Review Group Effort (SURGE) study authors. Physical distancing, face masks, and eye protection to prevent person-to-person transmission of SARS-CoV-2 and COVID-19: a systematic review and meta-analysis. Lancet. 2020 Jun 27;395(10242):1973-1987. https://doi.org/10.1016/S0140-6736(20)31142-9

3
Li Y, Liang M, Gao L, et al. Face masks to prevent transmission of COVID-19: A systematic review and meta-analysis. Am J Infect Control. 2021 Jul;49(7):900-906. https://doi.org/10.1016/j.ajic.2020.12.007

4
Liang M, Gao L, Cheng C, et al. Efficacy of face mask in preventing respiratory virus transmission: A systematic review and meta- analysis. Travel Med Infect Dis. 2020 Jul-Aug;36:101751. https://doi.org/10.1016/j.tmaid.2020.101751

5
Hemmer CJ, Hufert F, Siewert S, et al. Protection From COVID-19–The Efficacy of Face Masks. Dtsch Arztebl Int. 2021 Feb 5;118(5):59-65. https://doi.org/10.3238/arztebl.m2021.0119

6
Candevir A, Üngör C, Çizmeci Şenel F, et al. How efficient are facial masks against COVID-19? Evaluating the mask use of various communities one year into the pandemic. Turk J Med Sci. 2021 Dec 17;51(SI-1):3238-3245. https://doi.org/10.3906/sag-2106-190

7
Tran TQ, Mostafa EM, Tawfik GM, et al. Efficacy of face masks against respiratory infectious diseases: a systematic review and network analysis of randomized-controlled trials. J Breath Res. 2021 Sep 13;15(4). https://doi.org/10.1088/1752-7163/ac1ea5

8
Bartoszko JJ, Farooqi MAM, Alhazzani W, et al. Medical masks vs N95 respirators for preventing COVID-19 in healthcare workers: A systematic review and meta-analysis of randomized trials. Influenza Other Respir Viruses. 2020 Jul;14(4):365-373. https://doi.org/10.1111/irv.12745

9
MacIntyre CR, Chughtai AA. A rapid systematic review of the efficacy of face masks and respirators against coronaviruses and other respiratory transmissible viruses for the community, healthcare workers and sick patients. Int J Nurs Stud. 2020 Aug;108:103629. https://doi.org/10.1016/j.ijnurstu.2020.103629

10
Bundgaard H, Bundgaard JS, Raaschou-Pedersen DET, et al. Face masks for the prevention of COVID-19 - Rationale and design of the randomised controlled trial DANMASK-19. Dan Med J. 2020 Aug 18;67(9):A05200363.

11
Long Y, Hu T, Liu L, et al. Effectiveness of N95 respirators versus surgical masks against influenza: A systematic review and meta-analysis. J Evid Based Med. 2020 May;13(2):93-101. https://doi.org/10.1111/jebm.12381

12
Chou R, Dana T, Jungbauer R, et al. Masks for Prevention of Respiratory Virus Infections, Including SARS-CoV-2, in Health Care and Community Settings: A Living Rapid Review. Ann Intern Med. 2020 Oct 6;173(7):542-555. https://doi.org/10.7326/M20-3213

13
Nanda A, Hung I, Kwong A, et al. Efficacy of surgical masks or cloth masks in the prevention of viral transmission: Systematic review, meta-analysis, and proposal for future trial. J Evid Based Med. 2021 May;14(2):97-111. https://doi.org/10.1111/jebm.12424

14
Abboah-Offei M, Salifu Y, Adewale B, et al. A rapid review of the use of face mask in preventing the spread of COVID-19. Int J Nurs Stud Adv. 2021 Nov;3:100013. https://doi.org/10.1016/j.ijnsa.2020.100013

15
Coclite D, Napoletano A, Gianola S, et al. Face Mask Use in the Community for Reducing the Spread of COVID-19: A Systematic Review. Front Med (Lausanne). 2021 Jan 12;7:594269. https://doi.org/10.3389/fmed.2020.594269

16
Kim MS, Seong D, Li H, Chung SK, et al. Comparative effectiveness of N95, surgical or medical, and non-medical facemasks in protection against respiratory virus infection: A systematic review and network meta-analysis. Rev Med Virol. 2022 Sep;32(5):e2336. https://doi.org/10.1002/rmv.2336

17
Loeb M, Bartholomew A, Hashmi M, et al. Medical Masks Versus N95 Respirators for Preventing COVID-19 Among Health Care Workers: A Randomized Trial. Ann Intern Med. 2022 Dec;175(12):1629-1638. https://doi.org/10.7326/M22-1966

18
Xiao J, Shiu EYC, Gao H, et al. Nonpharmaceutical Measures for Pandemic Influenza in Nonhealthcare Settings-Personal Protective and Environmental Measures. Emerg Infect Dis. 2020 May;26(5):967-975. https://doi.org/10.3201/eid2605.190994

19
Daoud AK, Hall JK, Petrick H, et al. The Potential for Cloth Masks to Protect Health Care Clinicians From SARS-CoV-2: A Rapid Review. Ann Fam Med. 2021 Jan-Feb;19(1):55-62. https://doi.org/10.1370/afm.2640

20
Muller SM. Masks, mechanisms and Covid-19: the limitations of randomized trials in pandemic policymaking. Hist Philos Life Sci. 2021 Mar 25;43(2):43. https://doi.org/10.1007/s40656-021-00403-9

21
Li Y, Wei Z, Zhang J, et al. Wearing masks to reduce the spread of respiratory viruses: a systematic evidence mapping. Ann Transl Med. 2021 May;9(9):811. https://doi.org/10.21037/atm-20-6745

22
Pearce N, Vandenbroucke JP. Arguments about face masks and Covid-19 reflect broader methodologic debates within medical science. Eur J Epidemiol. 2021 Feb;36(2):143-147. https://doi.org/10.1007/s10654-021-00735-7

23
Fortaleza CR, Souza LDR, Rúgolo JM, et al. COVID-19: What we talk about when we talk about masks. Rev Soc Bras Med Trop. 2020 Nov 6;53:e20200527. https://doi.org/10.1590/0037-8682-0527-2020

24
Laine C, Goodman SN, Guallar E. The Role of Masks in Mitigating the SARS- CoV-2 Pandemic: Another Piece of the Puzzle. Ann Intern Med. 2021 Mar;174(3):419-420. https://doi.org/10.7326/M20-7448

25
Brainard J, Jones NR, Lake IR, et al. Community use of face masks and similar barriers to prevent respiratory illness such as COVID-19: a rapid scoping review. Euro Surveill. 2020 Dec;25(49):2000725. https://doi.org/10.2807/1560-7917.ES.2020.25.49.2000725

26
Aggarwal N, Dwarakanathan V, Gautam N, et al. Facemasks for prevention of viral respiratory infections in community settings: A systematic review and meta-analysis. Indian J Public Health. 2020 Jun;64(Supplement):S192-S200. https://doi.org/10.4103/ijph.IJPH_470_20

27
Baier M, Knobloch MJ, Osman F, et al. Effectiveness of Mask-Wearing on Respiratory Illness Transmission in Community Settings: A Rapid Review. Disaster Med Public Health Prep. 2022 Mar 7:1-8. https://doi.org/10.1017/dmp.2021.369

28
Rowan NJ, Moral RA. Disposable face masks and reusable face coverings as non-pharmaceutical interventions (NPIs) to prevent transmission of SARS-CoV-2 variants that cause coronavirus disease (COVID-19): Role of new sustainable NPI design innovations and predictive mathematical modelling. Sci Total Environ. 2021 Jun 10;772:145530. https://doi.org/10.1016/j.scitotenv.2021.145530

29
Hirt J, Janiaud P, Hemkens LG. Randomized trials on non-pharmaceutical interventions for COVID-19: a scoping review. BMJ Evid Based Med. 2022 Dec;27(6):334-344. https://doi.org/10.1136/bmjebm-2021-111825

30
Chen Y, Wang Y, Quan N, et al. Associations Between Wearing Masks and Respiratory Viral Infections: A Meta-Analysis and Systematic Review. Front Public Health. 2022 Apr 27;10:874693. https://doi.org/10.3389/fpubh.2022.874693

31
Lehnert B, Herold J, Blaurock M, et al. Reliability of the Acoustic Voice Quality Index AVQI and the Acoustic Breathiness Index (ABI) when wearing CoViD-19 protective masks. Eur Arch Otorhinolaryngol. 2022 Sep;279(9):4617-4621. https://doi.org/10.1007/s00405-022-07417-4

32
Dugré N, Ton J, Perry D, et al. Masks for prevention of viral respiratory infections among health care workers and the public: PEER umbrella systematic review. Can Fam Physician. 2020 Jul;66(7):509-517.

33
Wang H, Chen MB, Cui WY, et al. The efficacy of masks for influenza-like illness in the community: A protocol for systematic review and meta-analysis. Medicine (Baltimore). 2020 Jun 5;99(23):e20525. https://doi.org/10.1097/MD.0000000000020525

34
Yang HJ, Yoon H, Kang SY, et al. Respiratory Protection Effect of Ear-loop-type KF94 Masks according to the Wearing Method in COVID-19 Pandemic: A Randomized, Open-label Study. J Korean Med Sci. 2021 Jul 19;36(28):e209. https://doi.org/10.3346/jkms.2021.36.e209

35
Egan M, Acharya A, Sounderajah V, et al. Evaluating the effect of infographics on public recall, sentiment and willingness to use face masks during the COVID-19 pandemic: A randomised internet-based questionnaire study. BMC Public Health. 2021 Feb 17;21(1):367. https://doi.org/10.1186/s12889-021-10356-0

Karam C, Zeeni C, Yazbeck-Karam V, et al. Respiratory Adverse Events After LMA® Mask Removal in Children: A Randomized Trial Comparing Propofol to Sevoflurane. Anesth Analg. 2023 Jan 1;136(1):25-33. https://doi.org/10.1213/ANE.0000000000005945

37
Jackson AR, Hull JH, Hopker JG, et al. The impact of a heat and moisture exchange mask on respiratory symptoms and airway response to exercise in asthma. ERJ Open Res. 2020 Jun 22;6(2):00271-2019. https://doi.org/10.1183/23120541.00271-2019

38
Turkia M. The History of Methylprednisolone, Ascorbic Acid, Thiamine, and Heparin Protocol and I-MASK+ Ivermectin Protocol for COVID-19. Cureus. 2020 Dec 31;12(12):e12403. https://doi.org/10.7759/cureus.12403

39
Spang RP, Pieper K. The tiny effects of respiratory masks on physiological, subjective, and behavioral measures under mental load in a randomized controlled trial. Sci Rep. 2021 Oct 1;11(1):19601. https://doi.org/10.1038/s41598-021-99100-7

40
Brian MS, Carmichael RD, Berube FR, et al. The effects of a respiratory training mask on steady-state oxygen consumption at rest and during exercise. Physiol Int. 2022 May 16. https://doi.org/10.1556/2060.2022.00176

41
Tian L, Liu Y, Wei X, et al. A systematic review and meta-analysis of different mask ventilation schemes on management of general anesthesia in patients with respiratory failure. Ann Palliat Med. 2021 Nov;10(11):11587-11597. https://doi.org/10.21037/apm-21-2709

42
Regmi K, Lwin CM. Impact of non-pharmaceutical interventions for reducing transmission of COVID-19: a systematic review and meta-analysis protocol. BMJ Open. 2020 Oct 22;10(10):e041383. https://doi.org/10.1136/bmjopen-2020-041383

43
Li J, Qiu Y, Zhang Y, et al. Protective efficient comparisons among all kinds of respirators and masks for health-care workers against respiratory viruses: A PRISMA-compliant network meta-analysis. Medicine (Baltimore). 2021 Aug 27;100(34):e27026. https://doi.org/10.1097/MD.0000000000027026

44
Cirit Ekiz B, Köksal N, Tuna T, et al. Comparison of full-face and oronasal mask effectiveness in hypercapnic respiratory failure patients with non-invasive mechanical ventilation. Tuberk Toraks. 2022 Jun;70(2):157-165. English. https://doi.org/10.5578/tt.20229806

45
Thota B, Samantaray A, Vengamma B, et al. A randomised controlled trial of high-flow nasal oxygen versus non-rebreathing oxygen face mask therapy in acute hypoxaemic respiratory failure. Indian J Anaesth. 2022 Sep;66(9):644-650. https://doi.org/10.4103/ija.ija_507_22

46
Duong-Quy S, Ngo-Minh X, Tang-Le-Quynh T, et al. The use of exhaled nitric oxide and peak expiratory flow to demonstrate improved breathability and antimicrobial properties of novel face mask made with sustainable filter paper and *Folium Plectranthii amboinicii* oil: additional option for mask shortage during COVID-19 pandemic. Multidiscip Respir Med. 2020 Jun 1;15(1):664. https://doi.org/10.4081/mrm.2020.664

47
Toprak E, Bulut AN. The effect of mask use on maternal oxygen saturation in term pregnancies during the COVID-19 process. J Perinat Med. 2020 Nov 26;49(2):148-152. https://doi.org/10.1515/jpm-2020-0422

48
Ünal E, Özdemir A. The Effect of Correct Handwashing and Mask Wearing Training on Cardiac Patients' COVID-19 Fear and Anxiety. J Community Health Nurs. 2022 Apr-Jun;39(2):71-89. https://doi.org/10.1080/07370016.2022.2058201

49
Coelho SG, Segovia A, Anthony SJ, et al. Return to school and mask-wearing in class during the COVID-19 pandemic: Student perspectives from a school simulation study. Paediatr Child Health. 2022 May 5;27(Suppl 1):S15-S21. https://doi.org/10.1093/pch/pxab102

50
Felinska EA, Chen ZW, Fuchs TE, et al. Surgical Performance Is Not Negatively Impacted by Wearing a Commercial Full-Face Mask with Ad Hoc 3D-Printed Filter Connection as a Substitute for Personal Protective Equipment during the COVID-19 Pandemic: A Randomized Controlled Cross-Over Trial. J Clin Med. 2021 Feb 2;10(3):550. https://doi.org/10.3390/jcm10030550

51
Graham F. Daily briefing: Masks work against COVID, finds a huge randomized trial. Nature. 2021 Sep 2. https://doi.org/10.1038/d41586-021-02415-8

52
Ollila HM, Partinen M, Koskela J, et al. Face masks to prevent transmission of respiratory infections: Systematic review and meta-analysis of randomized controlled trials on face mask use. PLoS One. 2022 Dec 1;17(12):e0271517. https://doi.org/10.1371/journal.pone.0271517

53
Poncin W, Schalkwijk A, Vander Straeten C, et al. Impact of surgical mask on performance and cardiorespiratory responses to submaximal exercise in COVID-19 patients near hospital discharge: A randomized crossover trial. Clin Rehabil. 2022 Aug;36(8):1032-1041. https://doi.org/10.1177/02692155221097214

54
Paiva DN, Wagner LE, Dos Santos Marinho SE, et al. Effectiveness of an adapted diving mask (Owner mask) for non-invasive ventilation in the COVID-19 pandemic scenario: study protocol for a randomized clinical trial. Trials. 2022 Mar 18;23(1):218. https://doi.org/10.1186/s13063-022-06133-y

55
Benítez-Robaina S, Ramos-Macías Á, Borkoski-Barreiro S, et al. COVID-19 era: Hearing handicaps behind face mask use in hearing aid users. J Int Adv Otol. 2022 Nov;18(6):465-470. https://doi.org/10.5152/iao.2022.21578

56
Feng L, Zhang Q, Ruth N, et al. Compromised skin barrier induced by prolonged face mask usage during the COVID-19 pandemic and its remedy with proper moisturization. Skin Res Technol. 2022 Nov 25. https://doi.org/10.1111/srt.13214

57
Lin Q, Cai Y, Yu C, et al. Effects of Wearing Face Masks on Exercise Capacity and Ventilatory Anaerobic Threshold in Healthy Subjects During the COVID-19 Epidemic. Med Sci Monit. 2022 May 30;28:e936069. https://doi.org/10.12659/MSM.936069

58
Mohd Kamil MK, Yuen Yoong KP, Noor Azhar AM, et al. Non-rebreather mask and low-flow nasal cannula vs high-flow nasal cannula in severe COVID-19 pneumonia in the emergency department. Am J Emerg Med. 2023 Jan;63:86-93. https://doi.org/10.1016/j.ajem.2022.10.029

Nazir N, Saxena A. The effectiveness of high-flow nasal cannula and standard non-rebreathing mask for oxygen therapy in moderate category COVID-19 pneumonia: Randomised controlled trial. Afr J Thorac Crit Care Med. 2022 May 5;28(1):10.7196/AJTCCM.2022.v28i1.206. https://doi.org/10.7196/AJTCCM.2022.v28i1.206

60
Chou R. Comparative Effectiveness of Mask Type in Preventing SARS-CoV-2 in Health Care Workers: Uncertainty Persists. Ann Intern Med. 2022 Dec;175(12):1763-1764. https://doi.org/10.7326/M22-3219

61
Heidinger A, Falb T, Werkl P, et al. The Impact of Tape Sealing Face Masks on Visual Field Scores in the Era of COVID-19: A Randomized Cross-over Study. J Glaucoma. 2021 Oct 1;30(10):878-881. https://doi.org/10.1097/IJG.0000000000001922

62
Wang MX, Gwee SXW, Chua PEY, et al. Effectiveness of Surgical Face Masks in Reducing Acute Respiratory Infections in Non-Healthcare Settings: A Systematic Review and Meta-Analysis. Front Med (Lausanne). 2020 Sep 25;7:564280. https://doi.org/10.3389/fmed.2020.564280

63
Dost B, Kömürcü Ö, Bilgin S, Dökmeci H, et al. Investigating the Effects of Protective Face Masks on the Respiratory Parameters of Children in the Postanesthesia Care Unit During the COVID-19 Pandemic. J Perianesth Nurs. 2022 Feb;37(1):94-99. https://doi.org/10.1016/j.jopan.2021.02.004

64
Bánfai B, Musch J, Betlehem J, et al. How effective are chest compressions when wearing mask? A randomised simulation study among first-year health care students during the COVID-19 pandemic. BMC Emerg Med. 2022 May 8;22(1):82. https://doi.org/10.1186/s12873-022-00636-2

65
Boyle KG, Napoleone G, Ramsook AH, et al. Effects of the Elevation Training Mask® 2.0 on dyspnea and respiratory muscle mechanics, electromyography, and fatigue during exhaustive cycling in healthy humans. J Sci Med Sport. 2022 Feb;25(2):167-172. https://doi.org/10.1016/j.jsams.2021.08.022

66
Femi-Abodunde A, Olinger K, Burke LMB, et al. Radiology Dictation Errors with COVID-19 Protective Equipment: Does Wearing a Surgical Mask Increase the Dictation Error Rate? J Digit Imaging. 2021 Oct;34(5):1294-1301. https://doi.org/10.1007/s10278-021-00502-w

67
Schultheis WG, Sharpe JE, Zhang Q, et al. Effect of Taping Face Masks on Quantitative Particle Counts Near the Eye: Implications for Intravitreal Injections in the COVID-19 Era. Am J Ophthalmol. 2021 May;225:166-171. https://doi.org/10.1016/j.ajo.2021.01.021

68
Science M, Caldeira-Kulbakas M, Parekh RS, et al., Back-to-School COVID-19 School Study Group. Effect of Wearing a Face Mask on Hand-to-Face Contact by Children in a Simulated School Environment: The Back-to-School COVID-19 Simulation Randomized Clinical Trial. JAMA Pediatr. 2022 Dec 1;176(12):1169-1175. https://doi.org/10.1001/jamapediatrics.2022.3833

69
Abbasi S, Siddiqui KM, Qamar-Ul-Hoda M. Adverse Respiratory Events After Removal of Laryngeal Mask Airway in Deep Anesthesia Versus Awake State in Children: A Randomized Trial. Cureus. 2022 Apr 19;14(4):e24296. https://doi.org/10.7759/cureus.24296

70

Saxena A, Nazir N, Pandey R, et al. Comparison of Effect of Non-invasive Ventilation Delivered by Helmet vs Face Mask in Patients with COVID-19 Infection: A Randomized Control Study. Indian J Crit Care Med. 2022 Mar;26(3):282-287. https://doi.org/10.5005/jp-journals-10071-24155

71
Al Ali RA, Gautam B, Miller MR, et al. Laryngeal Mask Airway for Surfactant Administration Versus Standard Treatment Methods in Preterm Neonates with Respiratory Distress Syndrome: A Systematic Review and Meta-analysis. Am J Perinatol. 2022 Oct;39(13):1433-1440. https://doi.org/10.1055/s-0041-1722953

72
Iezadi S, Azami-Aghdash S, Ghiasi A, et al. Effectiveness of the non-pharmaceutical public health interventions against COVID-19; a protocol of a systematic review and realist review. PLoS One. 2020 Sep 29;15(9):e0239554. doi: 10.1371/journal.pone.0239554. Update in: PLoS One. 2021 Nov 23;16(11):e0260371.

73
Pin-On P, Leurcharusmee P, Tanasungnuchit S, et al. Desflurane is not inferior to sevoflurane in the occurrence of adverse respiratory events during laryngeal mask airway anesthesia: a non-inferiority randomized double-blinded controlled study. Minerva Anestesiol. 2020 Jun;86(6):608-616. https://doi.org/10.23736/S0375-9393.20.14202-0

**Appendix C – Results for other meta-analyses or systematic reviews**

Table C1. Outcome measures (effect size and 95% confidence intervals) and p-values for 5 cluster-randomized control trials (base studies) included in Aggarwal et al. (2020) meta-analysis.

| Outcome measure | 1st Author Year | Effect size (95% CI) | p-value |
|---|---|---|---|
| Self-reported ILI | Aiello 2010 | $-0.33$ ($-0.64 - -0.02$) | 0.0369 |
| Lab-confirmed Influenza (LCI) | Aiello 2012 | $-0.16$ ($-0.63 - 0.31$) | 0.5046 |
| " | Cowling 2008 | $0.69$ ($-0.56 - 1.95$) | 0.2812 |
| Lab-confirmed viral infection (Influenza) | MacIntyre 2009 | $0.25$ ($-0.43 - 0.94$) | 0.4744 |
| Self-reported ILI | Suess 2012 | $-0.49$ ($-1.85 - 0.86$) | 0.4785 |

Table C2. Outcome measures (risk ratio and 95% confidence intervals) and p-values for 7 randomized control trials (base studies) included in Xiao et al. (2020) & WHO (2019) meta-analysis.

| Outcome measure | 1st Author Year | Risk ratio (95% CI) | p-value |
|---|---|---|---|
| Lab-confirmed Influenza (LCI)a | Aiello 2010 | 2.34 (0.56 − 9.72) | 0.5663 |
| " | Aiello 2012 | 0.71 (0.34 − 1.48) | 0.3187 |
| " | Baeasheed 2014 | 7.43 (0.33 − 169.47) | 0.8815 |
| " | Cowling 2008 | 1.12 (0.37 − 3.35) | 0.8746 |
| " | Macintyre 2009 | 3.19 (0.13 − 77.36) | 0.9115 |
| " | Macintyre 2016 | 0.33 (0.01 − 7.96) | 0.7411 |
| " | Suess 2012 | 0.38 (0.38 − 0.89) | 0.0009 |

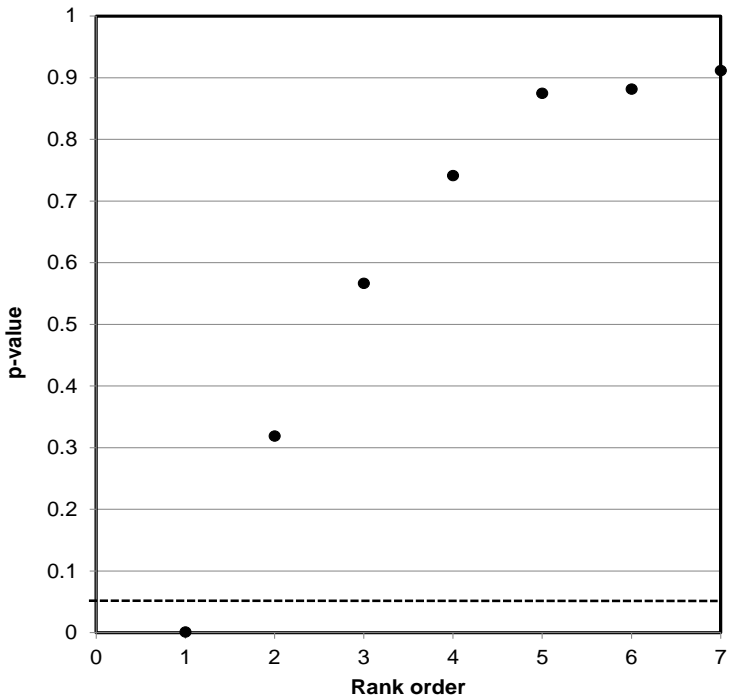

Figure C1. p-value plot for 7 randomized control trials (base studies) included in Xiao et al. (2020) & WHO (2019) meta-analysis.

Table C3. Outcome measure (risk ratio and 95% confidence intervals) and p-values for 7 randomized control trials (base studies) included in Nanda et al. (2021) meta-analysis.

| Outcome measure | 1st Author Year | Risk ratio (95% CI) | p-value |
|---|---|---|---|
| Lab-confirmed virus (Influenza) | Aiello 2010 | 0.99 (0.98 − 1.01) | 0.192629 |
| " | Aiello 2012 | 1.01 (0.99 − 1.04) | 0.436782 |
| " | Baeasheed 2014 | 0.92 (0.81 − 1.05) | 0.209509 |
| " | Cowling 2008 | 0.99 (0.92 −1.07) | 0.806279 |
| " | Macintyre 2009 | 0.97 (0.91 − 1.03) | 0.340394 |
| " | Macintyre 2016 | 1.01 (1.00 − 1.02) | 0.048497 |
| " | Suess 2012 | 1.19 (1.03 − 1.37) | 0.016719 |

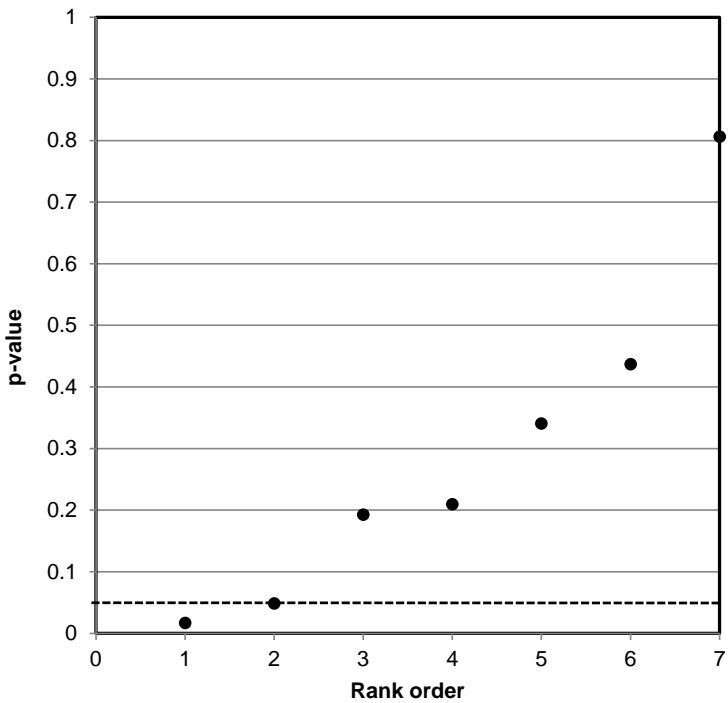

Figure C2. p-value plot for 7 randomized control trials (base studies) included in Nanda et al. (2021) meta-analysis.

Table C4. Outcome measure (odds ratio and 95% confidence intervals) and p-values for 7 randomized control trials (base studies) included in Kim et al. (2022) meta-analysis.

| Outcome measure | 1st Author Year | Odds ratio (95% CI) | p-value |
|---|---|---|---|
| LCI | Aiello 2012 | 0.7 (0.33 − 1.5) | 0.3148 |
| " | Alfelali 2020 | 1.16 (0.55 − 2.48) | 0.7452 |
| Lab-confirmed COVID | Bundgaard 2020 | 0.82 (0.55 − 1.23) | 0.2294 |
| LCI | Cowling 2008 | 1.16 (0.31 − 4.34) | 0.8763 |
| " | MacIntyre 2011 | 0.52 (0.13 − 2.09) | 0.3371 |
| " | MacIntyre 2009 | 4.96 (0.26 − 92.99) | 0.8671 |
| " | Suess 2012 | 0.32 (0.12 − 0.84) | 0.0002 |

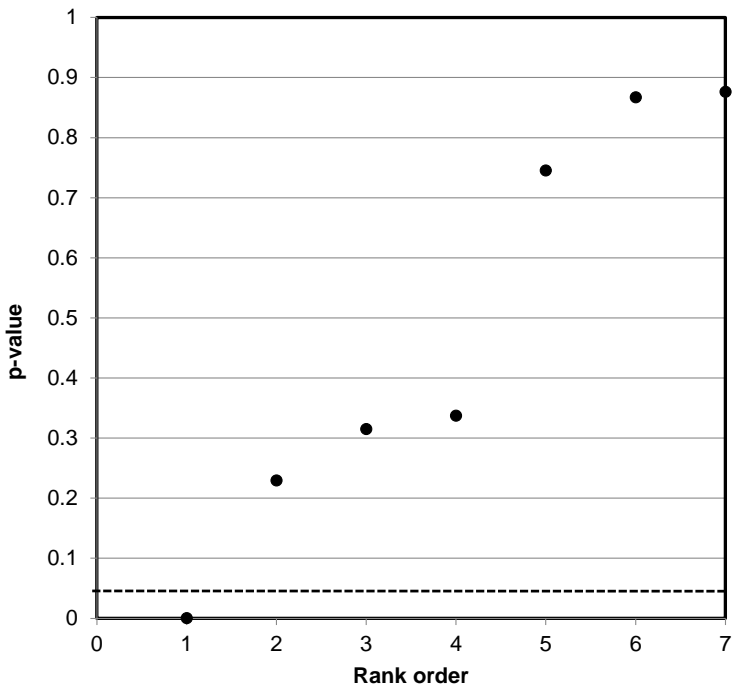

Figure C3. p-value plot for 7 randomized control trials (base studies) included in Kim et al. (2021) meta-analysis.

Table C5. Outcome measure (odds ratio and 95% confidence intervals) and p-values for 8 randomized control trials (base studies) included in Ollila et al. (2022) meta-analysis.

| Outcome measure | 1st Author Year | Odds ratio (95% CI) | p-value |
|---|---|---|---|
| Self-reported viral symptoms* | Barasheed 2014 | 0.393 (0.161 − 0.959) | |
| Self-reported viral symptoms* | Aiello 2010 | 0.709 (0.552 − 0.910) | |
| Unknown[o] | Aiello 2012 | 0.725 (0.497 − 1.058) | |
| Lab-reported COVID infection | Bundgaard 2020 | 0.815 (0.542 − 1.226) | |
| Self-reported ILI symptoms | Aelami 2015 | 0.874 (0.644 − 1.187) | |
| Self-reported COVID symptoms*[+] | Abaluck 2021 | 0.908 (0.829 − 0.995) | |
| Self-reported ARI symptoms | Abdin 2005 | 0.970 (0.733 − 1.284) | |
| Clinical confirmed respiratory infection | Alfelali 2020 | 1.089 (0.828 − 1.277) | |

* Laboratory-confirmed measures did not show a difference between mask and control groups.

[o] Unable to establish what statistics were used from review of base study article.

[+] The Chikina et al. (2022) re-analysis states that all of the outcomes in the study are based on self-reporting of symptoms.

Note: Ollila et al. do not state anywhere in their study which outcome measures were used.

Table C6. Outcome measure (p-values) for16 randomized control trials (base studies) included in Liu et al. (2021) systematic review.

| Base study | Intervention | Control group | Outcomes | P-value | Comments |
|---|---|---|---|---|---|
| Aiello et al. (2010) [U. Mich. Dorms] | medical mask (MM) | no MM | influenza-like illness (ILI); lab (PCR) confirmed influenza infection | 0.25 | for ILI; [*note: PCR results stated as non-significant, no data provided to estimate p-value, ILI data not used for p-value plot*] |
| Aiello et al. (2012) [U. Mich. dorms] | MM | no MM | ILI; lab (RT-PCR) confirmed influenza infection | 0.52 0.42 0.72 **0.69** | for ILI before adjustments for covariates; for ILI after adjustments for covariates; for RT-PCR before adjustments for covariates; for RT-PCR after adjustments for covariates |
| Abdin et al. (2005) [Hajj pilgrims] | MM | no MM | acute respiratory infection | 0.84 | for acute respiratory infection; (OR 0.97, 95% CI 0.73−1.28); p-value estimated; [*note: not used for p-value plot as base study unavailable to review*] |
| Barasheed et al. (2014) [Hajj pilgrims] | MM | no MM | ILI; lab (swab testing) confirmed virus infection | 0.04 **0.90** 0.90 | for ILI; for lab confirmed Influenza A virus infection; for lab confirmed Influenza B virus infection |
| Alfelali et al. (2020) [Hajj Pilgrims] | MM | no MM | respiratory virus infections (RVIs); lab-confirmed RVIs | 0.18 0.40 0.26 0.06 | for RVIs (intention-to-treat analysis); for lab-confirmed RVIs (intention-to-treat analysis); for RVIs (per-protocol analysis); for lab-confirmed RVIs (per-protocol analysis) [*note: not used for p-value plot as viruses included rhinovirus, Influenza viruses, parainfluenza viruses*] |
| Canini et al. (2010) [households in France] | MM | no MM | ILI – positive rapid Influenza A test | **1.00** | for difference in ILI between groups |
| Macintyre et al. (2009) [households in Australia] | MM | no MM | ILI; lab-confirmed total virus infections (VIs) | 0.50 0.46 0.32 | for ILI (by house); for ILI (by individual); for lab-confirmed total VIs; [*note: not used for p-value plot as viruses included Influenza A and B, respiratory syncytial virus, adenovirus, parainfluenza viruses (PIV) types 1–3, coronaviruses, human metapneumovirus, enteroviruses, rhinoviruses*] |
| Macintyre et al. (2016) [households in China] | MM | no MM | ILI; clinical respiratory illness (CRI); lab-confirmed viral illnesses | 0.34 0.44 **0.98** | for ILI; for clinical respiratory illness (CRI); for lab-confirmed viral illnesses; [p-values estimated] |
| Simmerman et al. (2011) [households in Thailand] | MM + hand washing | No intervention (MM or hand washing) | lab-confirmed Influenza (by RT-PCR or serology) | **0.525** | for lab-confirmed Influenza |

Note: p-values ***italicized & bolded*** used for p-value plot.

Table C6. Outcome measure (p-values) for16 randomized control trials (base studies) included in Liu et al. (2021) systematic review (con't).

| Base study | Intervention | Control | Outcomes | P-value | Comments |
|---|---|---|---|---|---|
| Cowling et al. (2008) [households in Hong Kong] | MM | no MM | lab-confirmed Influenza; clinical Influenza definition 1; clinical Influenza definition 2; clinical Influenza definition 3 | *0.99* 1.00 0.97 *0.52* | for lab-confirmed Influenza; for clinical Influenza definition 1; for clinical Influenza definition 2; for clinical Influenza definition 3 |
| Cowling et al. (2009) [households in Hong Kong] | MM + hand hygiene | No intervention (MM or hand hygiene) | Influenza A + B virus infection confirmed by RT-PCR; clinical diagnosis after 7 days (2 definitions) | *0.48* 0.37 *0.26* | for lab (RT-PCR)-confirmed Influenza (OR 0.77, 95% CI 0.38–1.55); for clinical Influenza definition 1 (OR 1.25, 95% CI 0.79–1.98); for clinical Influenza definition 2 (OR 1.68, 95% CI 0.68–4.15); [p-values estimated] |
| Suess et al. (2012) [households in Germany | MM | no MM | lab-confirmed (RT-PCR) for Influenza; clinical ILI | *0.10* *0.30* | for lab-confirmed (RT-PCR); for clinical ILI |
| Larson et al. (2010) [Hispanic households in New York City] | MM + hand sanitizer + education | Education only (i.e., no MM + hand sanitizer) | Influenza (A or B) confirmatory testing by culture or RT-PCR; ILI (CDC definition); viral upper respiratory infections | 0.893 0.61 0.194 | for Influenza (RT-PCR lab-confirmed); for ILI; for viral upper respiratory infections |
| Jacobs et al. (2009) [hospital workers in Japan] | MM | no MM | presence of a cold based on a previously validated measure of self-reported symptoms | 0.81 | for presence of a cold; 32 health care workers completed the study; 8 symptoms recorded daily [*note: not used for p-value plot as base study unavailable to review*] |
| Bundgaard et al. (2021) [adult community members in Denmark] | MM | no MM | SARS-CoV-2 infection at 1 month by: IgM antibody testing; IgG antibody testing; RT-PCR, or healthcare diagnosed | 0.35 *0.58* *0.80* – *0.23* | for SARS-CoV-2 infection: main trial measurement end point (OR 0.82, 95% CI 0.54–1.23); +ve IgM antibody test result (OR 0.87, 95% CI 0.54–1.41); +ve IgG antibody test result (OR 1.07, 95% CI 0.66–1.75); RT-PCR positivity (n/a); healthcare diagnosed (OR 0.52, 95% CI 0.18–1.53); [p-values estimated] |
| Abaluck et al. (2021) [cluster randomized communities in Bangladesh] | MM | no MM | reduction in symptomatic SARS-CoV-2 seroprevalence | 0.066 *0.062* | reduced symptomatic SARS-CoV-2 seroprevalence, 2 results are given: n=200 blood samples, symptomatic seroprevalence adjusted prevalence ratio = 0.89 [0.78, 1.00]; n=10,790 blood samples, symptomatic seroprevalence adjusted prevalence ratio = 0.91 [0.82, 1.00]; [p-values estimated; note: results for non-medical, cloth masks excluded] |

Note: p-values ***italicized & bolded*** used for p-value plot.